%% file: ggHH.tex
\documentclass[a4paper,11pt]{article}
\pdfoutput=1

\usepackage{amsmath}
\usepackage{jheppub}
\usepackage[utf8]{inputenc}
\usepackage[T1]{fontenc}
\usepackage[kerning,spacing]{microtype}
\usepackage{graphicx}
\usepackage{amsbsy,amssymb}
\usepackage{mathtools}
\usepackage{xspace}
\usepackage{dcolumn}
\usepackage{bm}
\usepackage{slashed}
\usepackage{color}
\usepackage[font=small]{caption}
\usepackage{subcaption}
\usepackage{booktabs}

\newcommand{\mhh}{\ensuremath{m_{\mathrm{hh}}}\xspace}
\newcommand{\pthh}{\ensuremath{p_{T}^{\mathrm{hh}}}\xspace}
\newcommand{\ptj}{p_{T}^{\mathrm{j}_1}}
\newcommand{\pth}{p_{T}^{\mathrm{h}}}
\newcommand{\pthl}{p_{T}^{\mathrm{h}_1}}
\newcommand{\pths}{p_{T}^{\mathrm{h}_2}}
\newcommand{\dphihh}{\Delta\Phi^{\mathrm{hh}}}
\newcommand{\drhh}{\Delta R^{\mathrm{hh}}}
\newcommand{\kt}{k_{\scriptscriptstyle\rm T}}
\newcommand{\GeV}{\ensuremath{\mathrm{GeV}}\xspace}
\newcommand{\TeV}{\ensuremath{\mathrm{TeV}}\xspace}

\newcommand{\eps}{\epsilon}

\newcommand{\powheg}{{\tt POWHEG}\xspace}
\newcommand{\powhegbox}{{\tt POWHEG-BOX}}
\newcommand{\mg}{{\tt MG5\_aMC@NLO}\xspace}
\newcommand{\gosam}{{\tt GoSam}}
\newcommand{\qgraf}{{\tt QGRAF}}
\newcommand{\form}{{\tt FORM}}

\newcommand{\ninja}{{\tt Ninja}}
\newcommand{\spinney}{{\tt Spinney}}
\newcommand{\avholo}{{\tt OneLOop}}
\newcommand{\fastjet}{{\tt Fastjet}\xspace}
\newcommand{\pythia}{{\tt Pythia\,8}\xspace}
\newcommand{\pythiaold}{{\tt Pythia\,6}\xspace}

\newcommand{\ftapprox}{FT$_{\mathrm{approx}}$\xspace}
\newcommand{\hdamp}{{\tt hdamp}\xspace}

\title{NLO predictions for Higgs boson pair production with full top quark mass dependence matched to parton showers}
\preprint{\vbox{\hbox{CERN-TH-2017-069}\hbox{MPP-2017-41}\hbox{NIKHEF-2017-020}}}
\author[a]{G.~Heinrich,}
\author[a]{S.~P.~Jones,}
\author[a]{M.~Kerner,}
\author[b]{G.~Luisoni,}
\author[c]{E.~Vryonidou}
\affiliation[a]{Max Planck Institute for Physics, F\"ohringer Ring 6, 80805 M\"unchen, Germany}
\affiliation[b]{Theoretical Physics Department, CERN, Geneva, Switzerland}
\affiliation[c]{Nikhef, Science Park 105, 1098 XG, Amsterdam, The Netherlands}

\emailAdd{gudrun@mpp.mpg.de, sjones@mpp.mpg.de, kerner@mpp.mpg.de,
  gionata.luisoni@cern.ch, eleniv@nikhef.nl}

\keywords{QCD, Higgs, NLO, Parton shower}

\abstract{%
  We present the first combination of NLO QCD matrix elements for di-Higgs
  production, retaining the full top quark mass dependence, with a
  parton shower.  Results are provided within both the {\tt POWHEG-BOX}
  and {\tt MadGraph5\_aMC@NLO} Monte Carlo frameworks.  We assess in detail
  the theoretical uncertainties and provide differential
  results. We find that, as expected, the shower effects are relatively large for observables like the transverse momentum of the Higgs boson pair,  
  which are sensitive to extra radiation. However, these shower effects are still much smaller than 
  the differences between the Born-improved HEFT approximation and the full NLO calculation in the tails of the distributions.}

\begin{document}

\maketitle

\section{Introduction}\label{sec:intro}

\input{intro}

\section{Details of the calculation}\label{sec:calculation}

\input{calculation}

\section{Results}\label{sec:results}

\input{results}

\section{Conclusions}

We have presented the combination of the full NLO prediction for Higgs
boson pair production, including the top quark mass dependence at two
loops, with a parton shower.  This has been implemented within two
frameworks, {\tt POWHEG-BOX} and {\tt MG5\_aMC@NLO}, using the same
{\tt Pythia 8.2} shower in both cases.  Individual phase-space points
of the two-loop amplitude, which depends only on the two independent
kinematic invariants $\hat{s}$ and $\hat{t}$ once the top-quark and
Higgs boson masses are fixed, have been used to create a grid and
combined with an interpolation framework, such that a value for the
amplitude can be obtained at any phase space point without
re-evaluating the loop integrals.

We find that the impact of the parton shower on the transverse
momentum distribution of one Higgs boson, $\pth$, is quite small and
that the features of the various approximations that have appeared
previously in the literature are preserved by the shower.

The impact of the shower on the $\pthh$, $\Delta\Phi^{hh}$ and $\Delta
R^{hh}$ distributions is fairly large, as these are the distributions
where the tail is predicted at the first non-trivial order in the
fixed order calculation.  In the tail of the $\pthh$ distribution,
around $\pthh\sim 400$\,GeV, the showered NLO results are larger than
the fixed order results by more than a factor of two, within both {\tt
  POWHEG} and {\tt MG5\_aMC@NLO}.  This feature is also present if
{\tt Pythia 6} is used instead of {\tt Pythia 8}, and if we vary the
shower starting scale in \mg.  However, the differences due to the
shower in the $\pthh$ distribution are still much smaller than the
discrepancy between the showered full calculation and the showered
Born-improved HEFT approximation, the latter overshooting the full
result by an order of magnitude around $\pthh\sim 400$\,GeV, worsening
towards higher $\pthh$ values.  As expected, the \ftapprox results,
which include the full mass dependence in the real radiation, behave
very similar to the full calculation in the (real radiation dominated)
tails of the distributions like $\pthh$.

In summary, we observe that the inclusion of the full mass dependence
in general has a more important impact on the distributions relevant
to Higgs boson pair production than effects coming from different
shower matching schemes, variations of the shower starting scales,
different parton showers or different PDFs.  A detailed study of
hadronisation effects and Higgs boson decays will be performed in a
subsequent publication.

The \powheg{} version of the code for Higgs boson pair production
developed for this work is 
publicly available in the \powhegbox{} {\tt V2} package, under the
{\tt User-Processes-V2/ggHH/} directory, and will become available
also in the newer \powhegbox{} {\tt RES} version in the {\tt
  User-Processes-RES/ggHH/} folder.  All the information can be found
at the web page {\tt http://powhegbox.mib.infn.it}.
The implementation in \mg is not part of the public release yet, 
but the customised code can be obtained by contacting the authors. 

We hope that making two-loop results available in the form of a grid
included in public Monte Carlo programs, as done in this work, will
open the door to further developments in this direction.

\section*{Acknowledgements}
We would like to thank our collaborators Sophia Borowka, Nicolas
Greiner, Johannes Schlenk and Tom Zirke for their contribution to the
double Higgs fixed order NLO project and Simone Alioli for interesting
discussions and suggestions concerning the {\tt POWHEG-BOX}.  We also
thank Javier Mazzitelli for providing us the NNLO HEFT data for
comparisons. We also would like to thank Fabio Maltoni and Paolo
Torrielli for useful discussions and Carlo Oleari for comments on the
manuscript.  G.L. thanks the Max Planck Institute for Physics for
hospitality during several phases of this work.  This research was
supported in part by the Research Executive Agency (REA) of the
European Union under the Grant Agreement PITN-GA2012316704
(HiggsTools).  E.V. acknowledges support from the Netherlands
Foundation for Fundamental Research of Matter (FOM) under project
number 15PR3224.  We gratefully acknowledge support and resources
provided by the Max Planck Computing and Data Facility (MPCDF).

\bibliographystyle{JHEP}
\bibliography{references}
\end{document}

%% file: intro.tex
Exploring the Higgs sector is one of the major goals for the next
phases of LHC experiments.  In particular, the form of the Higgs
potential as predicted by the Standard Model (SM) needs to be
confirmed.  While one important parameter of the potential, the Higgs
boson mass, has been measured already to an impressive accuracy, the
Higgs boson self-coupling is still only very weakly constrained.  The
latter can be measured for example via Higgs boson pair production in
gluon fusion, which is the dominant production mechanism of Higgs
boson pairs.  However, the cross section is about $1000$ times
smaller than that for single Higgs production, which makes the measurement 
very challenging even with the high luminosity upgrade of the LHC.
This fact on the other hand makes this channel very interesting for
New Physics searches, as the delicate cancellations between different
contributions which happen in the SM are altered in most New Physics
models, leading to potentially large effects.

At the LHC, the decay channel $HH\to b\bar{b}\gamma\gamma$ has
so far led to the most stringent limit on the di-Higgs production
cross section of 
$\sigma/\sigma_{SM}\leq 19$ in CMS~\cite{CMS-PAS-HIG-17-008}, 
while the ATLAS collaboration
achieved the most restrictive upper bound of $\sigma/\sigma_{SM}\leq 29$
in the $b\bar{b}b\bar{b}$ decay channel~\cite{ATLAS-CONF-2016-049}.

A previous combination of
various decay channels measured in the ATLAS detector led to
$\sigma/\sigma_{SM}\leq 70$~\cite{Aad:2015xja}.  The CMS collaboration
also produced new limits for resonant and non-resonant Higgs
boson pair production in the $b\bar{b}VV$ channel~\cite{CMS-PAS-HIG-17-006}.


On the theory side, the leading order calculation of Higgs boson pair
production in gluon fusion, which proceeds via heavy quark loops, has
been performed in
Refs.~\cite{Eboli:1987dy,Glover:1987nx,Plehn:1996wb}. 
Higher order corrections were for a long time available only within the
Higgs Effective Field Theory (HEFT) approximation, where the NLO corrections are
calculated in the $m_t\to\infty$ limit, leading to point-like
effective couplings of gluons to Higgs bosons. In
Ref.~\cite{Dawson:1998py} NLO corrections were calculated in the
so-called ``Born-improved HEFT'' approximation, where the basic HEFT
result is rescaled by a factor $B_{FT}/B_{HEFT}$, $B_{FT}$
denoting the leading order matrix element squared in the full theory.


In Refs.~\cite{Frederix:2014hta,Maltoni:2014eza}, an approximation called
``\ftapprox'' was introduced, which contains the full top quark
mass dependence in the real radiation, while the virtual part is
calculated in the HEFT approximation and rescaled at the event level by
the re-weighting factor $B_{FT}/B_{HEFT}$.

In addition, the HEFT results at NLO and NNLO have been improved by an
expansion in $1/m_t^{2\rho}$ in
Refs.~\cite{Grigo:2013rya,Grigo:2014jma,Grigo:2015dia,Degrassi:2016vss},
with $\rho^{\rm{max}}=6$ at NLO, and $\rho^{\rm{max}}=2$ for the
soft-virtual part at NNLO ~\cite{Grigo:2015dia}.

The NNLO QCD corrections in the heavy top limit have been computed in
Refs.~\cite{deFlorian:2013uza,deFlorian:2013jea,Grigo:2014jma,deFlorian:2016uhr},
and they have been supplemented by an expansion in $1/m_t^2$ in
Ref.~\cite{Grigo:2015dia} and by resummation, at NLO+NNLL in
Ref.~\cite{Shao:2013bz} and at NNLO+NNLL in
Ref.~\cite{deFlorian:2015moa}, leading to K-factors of about 1.2
relative to the Born-improved HEFT result.

Very recently, the full NLO corrections, including the top quark mass
dependence also in the virtual two-loop amplitudes, have been
calculated~\cite{Borowka:2016ehy,Kerner:2016msj} and compared to
previous approximations for various
observables~\cite{Borowka:2016ypz}.  The full NLO calculation was
supplemented by NLL resummation in Ref.~\cite{Ferrera:2016prr}.


Numerous phenomenological studies of Higgs boson pair production have
been performed both within and beyond the
SM~\cite{Dolan:2012rv,Baglio:2012np,Goertz:2013kp,Gouzevitch:2013qca,Barr:2013tda,Dolan:2013rja,Barger:2013jfa,Li:2013flc,deLima:2014dta,Slawinska:2014vpa,Buschmann:2014sia,Azatov:2015oxa,Dicus:2015yva,Papaefstathiou:2015iba,Grober:2015cwa,Lu:2015jza,Dawson:2015oha,Ghezzi:2015vva,Dolan:2015zja,DallOsso:2015aia,Dawson:2015haa,Cao:2015oaa,Cao:2015oxx,Behr:2015oqq,Li:2016nrr,Kanemura:2016lkz,Agostini:2016vze,Grober:2016wmf,Banerjee:2016nzb,Huang:2017jws}.
Further, it also has been suggested recently to obtain constraints on the
Higgs boson self-coupling from electroweak corrections to single Higgs
boson production~\cite{Gorbahn:2016uoy,Degrassi:2016wml,Bizon:2016wgr}.

The studies of Higgs boson pair production mentioned above usually had at least one of the following drawbacks:
either they are based on leading order matrix elements, while
including the full top quark mass dependence, or the matrix elements
include higher orders in QCD but have been performed within the
infinite-top-mass approximation, which is known to fail at scales
where the top quark loops are
resolved~\cite{Baur:2002rb,Borowka:2016ehy,Borowka:2016ypz}.

Results for Higgs boson pair production merged to $HH+1$ jet matrix
elements at leading order, with full top and bottom quark mass
dependence, matched to a parton shower within {\tt HERWIG++}, have
been presented in Ref.~\cite{Maierhofer:2013sha}.
The ``\ftapprox''~\cite{Maltoni:2014eza} calculation includes the
matching of di-Higgs production to a parton
shower~\cite{Frederix:2014hta} keeping the full top quark mass
dependence in the real radiation, while the virtual part is calculated
in the Born-improved HEFT approximation.

In this paper, we present the first combination of the full NLO
calculation, including the full top quark mass dependence at two
loops, with a parton shower, within both the {\tt
POWHEG-BOX}~\cite{Frixione:2007vw,Alioli:2010xd} and the {\tt
MadGraph5\_aMC@NLO} framework~\cite{Alwall:2014hca,Hirschi:2015iia}.  This
allows us to compare the {\tt POWHEG}~\cite{Frixione:2007vw} and {\tt
MC@NLO}~\cite{Frixione:2002ik} matching schemes while using the same
{\tt Pythia 8}~\cite{Sjostrand:2007gs,Sjostrand:2014zea} shower in
both cases.  We also investigate the PDF and scale uncertainties and
calculate observables like the di-Higgs $p_T$ spectrum, $p_T^{hh}$, where fixed
order NLO calculations cannot give a satisfactory description at low
$p_T^{hh}$.  
Further, we discuss the possibility to infer the leading contribution of a full NNLO
calculation from the showered results, based on a comparison of 
the NNLO calculation in the HEFT approximation~\cite{deFlorian:2016uhr} with the showered results.

%% file: calculation.tex
In this section we present the details of the implementation of the
calculation within the \powhegbox{} Monte Carlo program.  The results
from \mg{} presented in the next sections are based on a similar
implementation. Both codes use the same grid for the virtual two-loop
amplitude discussed in Sec.~\ref{subsec:virtual}. Further details
about the calculation based on Born-improved HEFT and \ftapprox within
\mg{} already have been published
elsewhere~\cite{Frederix:2014hta,Maltoni:2014eza}.

In order to allow for comparisons and cross checks, we implemented
both the effective theory as well as the full SM amplitudes at
NLO. This allows to run the code in four different modes by changing
the flag {\tt mtdep} in the \powhegbox{} run card. The possible
choices and the corresponding calculation, as presented in the
previous section, are the following:
\begin{description}
 \item[{\tt mtdep=0}:]{computation using basic HEFT,}
 \item[{\tt mtdep=1}:]{computation using Born-improved HEFT,}
 \item[{\tt mtdep=2}:]{computation in the approximation \ftapprox (full
   mass dependence in the Born and in the real radiation, Born-improved HEFT
   for the virtual part),}
 \item[{\tt mtdep=3}:]{computation in the full SM.}
\end{description}
The corresponding modes are also available in \mg.

The leading order amplitude in the full theory and  all the
amplitudes in the HEFT were implemented analytically, whereas the
one-loop real radiation contribution and the two-loop virtual
amplitudes in the full SM rely on numerical or semi-numerical
codes. Since the virtual two-loop amplitudes in the full theory are
computed keeping the Higgs bosons on-shell, we assume a vanishing
Higgs boson width in all the modes listed above. Higgs boson decays
can be computed in the narrow width approximation by the parton
shower. In the next sections we give some more details about our
implementation. We will use the term ``HEFT approximation'', or simply
``HEFT'', for basic HEFT, while results in the Born-improved HEFT
will be denoted by ``B-i. HEFT''.

\subsection{Virtual two-loop amplitudes}\label{subsec:virtual}

\input{grid}

\subsection{Real radiation  and parton shower matching}

\input{shower}

%% file: grid.tex
For the virtual two-loop amplitudes, we have used the results of the
calculation presented in Refs.~\cite{Borowka:2016ehy,Borowka:2016ypz},
which is based on an extension of the program
\gosam{}~\cite{Cullen:2014yla} to two loops~\cite{Jones:2016bci},
using also {\sc Reduze}\,2~\cite{vonManteuffel:2012np} and {\sc
 SecDec}\,3~\cite{Borowka:2015mxa}.

The values for the Higgs boson and top quark masses have been set to
$m_h=125$\,GeV and $m_t=173$\,GeV, such that the two-loop amplitudes
only depend on two independent variables, the Mandelstam invariants
$\hat{s}$ and $\hat{t}$.  We have constructed a grid in these
variables together with an interpolation framework, such that an
external program can call the virtual two-loop amplitude at any phase space
point without having to do costly two-loop integrations.

\begin{figure}[t]
\centering
\begin{subfigure}{0.49\textwidth}
\includegraphics[width=\textwidth]{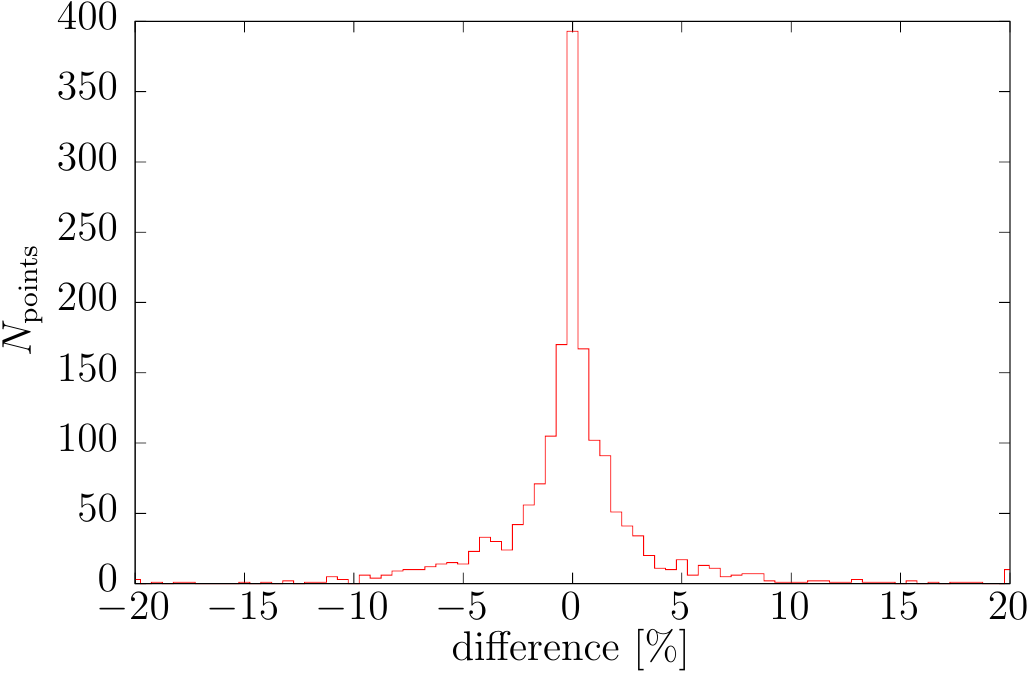}
\end{subfigure}
\begin{subfigure}{0.49\textwidth}
\includegraphics[width=\textwidth]{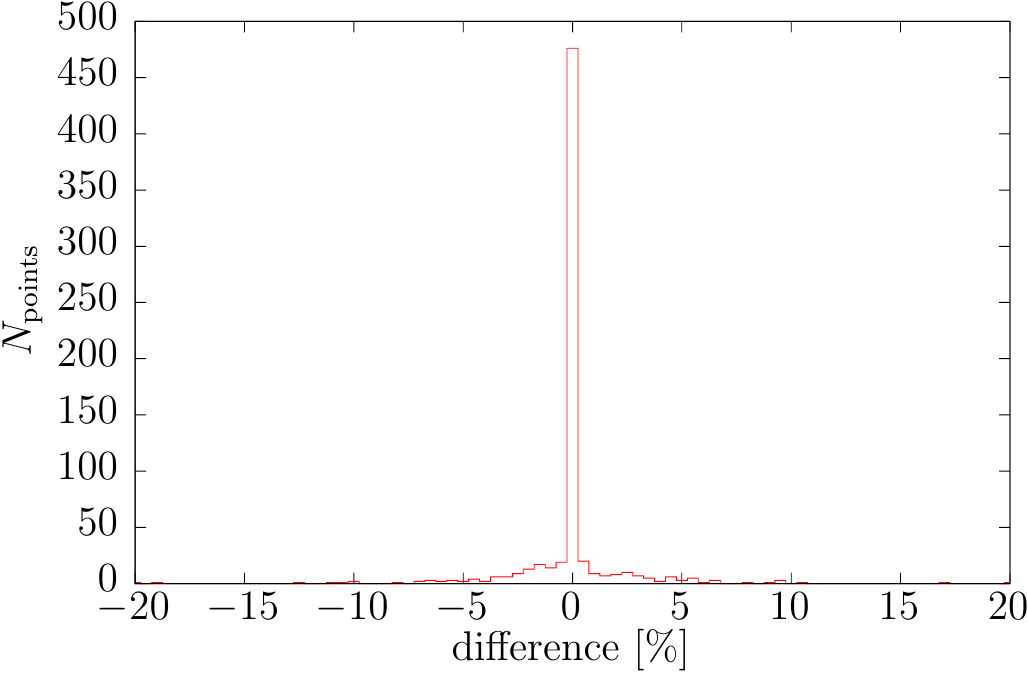}
\end{subfigure}
\caption{Closure test of the grid interpolation. The left~(right) plot
shows the relative difference of the grid results compared to a grid obtained 
from 50\%~(80\%) of the input data points, evaluated at the remaining data points.
Differences are defined as positive~(negative) if the full grid yields 
larger~(smaller) results.
The outermost bins contain all results with differences larger than 20\%.
\label{fig:closureTest}
}
\end{figure}

In more detail, we first transform the Mandelstam invariants $\hat{s}$
and $\hat{t}$ to new variables
\begin{align}
  x &= f(\beta(\hat{s})),\quad \textrm{with} \quad\beta=\left( 1-\frac{4 m_{h}^2}{\hat{s}} \right)^\frac{1}{2}\\
  c_\theta &= |\cos \theta\, | = \left|\frac{\hat{s}+2\hat{t}-2m_h^2}{\hat{s} \beta(\hat{s})}\right|,
  \label{eq:gridVariables}
\end{align}
where $f$ can, in principle, be any strictly increasing
function. Setting $f(\beta)$ according to the cumulative distribution
function of the phase space points used in our original calculation,
we obtain a nearly uniform distribution of these points in the
$(x,c_\theta)$ unit square.  Instead of a direct interpolation of the
phase space points, we chose to apply a two-step procedure: First, we
generate a regular grid with a fixed grid spacing in the variables $x$
and $c_\theta$, where we estimate the result at each grid point
applying a linear interpolation of our original results in the
vicinity of the specific grid point. In a second step, we apply
Clough-Tocher interpolation~\cite{CloughTocher} as implemented in the
python \textit{SciPy} package~\cite{scipy}.  Applying this procedure
reduces the size of interpolation artefacts, which we obtain due to
the numerical uncertainty of our two-loop results.
In Fig.~\ref{fig:closureTest} we test how omitting input data points
influences the results of the grid interpolation.
Removing 20\% of the input data points changes the interpolation results 
by less than 0.25\% for 70\% of the tested points. Differences larger than
5\% are obtained for 6\% of the results.

We should point out that the grid is constructed from a sample of
phase space points which is based on runs at
$\sqrt{s}=14$\,TeV. Therefore, even though the grid is not explicitly
dependent on the centre-of-mass energy, one should be aware of the
fact that for runs at e.g. 100\,TeV, the grid may not be reliable for
points with large $\hat s$ due to a lack of statistics in this region
upon construction.

In our original calculation~\cite{Borowka:2016ehy,Borowka:2016ypz}, we
used Catani-Seymour dipole subtraction~\cite{Catani:1997vz} for the
real radiation.  The finite combination of the renormalized virtual
amplitude $\mathcal V_b$ with the Catani-Seymour {\bf I}-operator can
be straightforwardly converted into the quantity ${\cal V}_{\rm{fin}}$
of Refs.~\cite{Frederix:2009yq,Alioli:2010xd}, defined by
\begin{eqnarray}
\mathcal{V}_{b}&=&\mathcal{N} \, \frac{\alpha_s}{2 \pi} \Bigg[ \frac{1}{\eps^2}\,a\,
\mathcal{B} +\frac{1}{\eps}\sum_{i,j} c_{ij}\,
\mathcal{B}_{ij}+\mathcal{V}_{\rm{fin}}\Bigg],
\label{eq:virtfin} \\
\mathcal{N} &=& \frac{(4 \pi)^{\epsilon}}{\Gamma (1 - \epsilon)} \left(
  \frac{\mu_r^2}{Q^2} \right)^{\epsilon}\;.
\end{eqnarray}
For this process the colour-correlated Born squared amplitudes
$\mathcal B_{12}$ and $\mathcal B_{21}$ are equal and we have $a=-2
C_A$ and $c_{12} = c_{21} = -\beta_0/2-C_A\ln\left(\mu_r^2/\hat{s}\right)$. In the \powhegbox{} and \mg frameworks
the arbitrary scale $Q$ is chosen as $\mu_r$.  Specifically, we obtain
\begin{eqnarray}\label{eq:vfin}
  \begin{split}
  \mathcal V_{\rm fin}(\mu_r)=&
  \frac{2\pi}{\alpha_s(\mu_r)}\Big(\mathcal V_b+\mathbf I \otimes  \mathcal B\Big)(\mu_r)\\
  &-\mathcal B(\mu_r) \Big(C_A \ln^2\Big(\frac{\mu_r^2}{\hat{s}}\Big)+\beta_0\ln\Big(\frac{\mu_r^2}{\hat{s}}\Big)+\beta_0+2 K_g-\frac{2 \pi^2}{3}C_A\Big).
  \label{eq:vfinConversion}
  \end{split}
\end{eqnarray}
The grid evaluates $\mathcal{V}_{\rm fin}$ at the scale
$\mu_0=\sqrt{\hat{s}}/2$ and the results for an arbitrary scale can be
obtained from the relation
\begin{align}
  \mathcal V_{\rm fin}(\mu_r) = \mathcal V_{\rm fin}(\mu_0)\cdot\frac{\mathcal B(\mu_r)}{\mathcal B(\mu_0)}+C_A \mathcal B(\mu_r)\left( \ln^2\left(\frac{\mu_0^2}{\hat{s}}\right)- \ln^2\left(\frac{\mu_r^2}{\hat{s}}\right) \right).
  \label{eq:vfinScaleDependence}
\end{align}

The Born amplitude $\mathcal{B}$ and the colour-correlated Born
amplitudes $\mathcal{B}_{ij}$ in \eqref{eq:virtfin} are evaluated in
$D=(4-2\eps)$ dimensions using conventional dimensional regularization
(CDR). As all formulas for the soft contributions and the collinear
remnants used in the \powhegbox{} and \mg are computed in the
$\overline{\rm{MS}}$ scheme, using CDR, constructing
$\mathcal{V}_{\rm{fin}}$ according to~(\ref{eq:vfin}) ensures the
treatment is consistent with that implemented in the Monte Carlo programs.

%% file: shower.tex
The real radiation matrix elements in the full SM were implemented
using the interface~\cite{Luisoni:2013cuh}
between \gosam~\cite{Cullen:2011ac,Cullen:2014yla} and
the \powhegbox~\cite{Frixione:2007vw,Alioli:2010xd}, modified
accordingly to compute the real corrections instead of the virtual
ones. The one-loop real amplitudes we generated with the new version
2.0 of \gosam{}~\cite{Cullen:2014yla}, that
uses \qgraf{}~\cite{Nogueira:1991ex}, \form~\cite{Kuipers:2012rf} and
\spinney{}~\cite{Cullen:2010jv} for the generation of the Feynman
diagrams, and offers a choice from {\tt
Samurai}~\cite{Mastrolia:2010nb,vanDeurzen:2013pja}, {\tt
golem95C}~\cite{Binoth:2008uq,Cullen:2011kv,Guillet:2013msa}
and \ninja{}~\cite{vanDeurzen:2013saa,Peraro:2014cba} for the
reduction.  At run time the amplitudes were computed using
\ninja{}~\cite{vanDeurzen:2013saa,Peraro:2014cba} and \avholo{}~\cite{vanHameren:2010cp}
for the evaluation of the scalar one-loop integrals.

In order to avoid numerical instabilities in the one-loop real matrix
elements in the limit where the additional parton becomes soft and/or
collinear, a technical cut $\pthh>10^{-3}$~\GeV has been
introduced. We carefully checked that the total cross section does not
change significantly when varying the cut value.


\medskip

Within \mg, the one-loop born and real amplitudes are computed using {\tt MadLoop}~\cite{Hirschi:2011pa}, which in turn
exploits {\tt CutTools}~\cite{Ossola:2007ax}, \ninja~\cite{Peraro:2014cba,Hirschi:2016mdz} or {\tt Collier}~\cite{Denner:2016kdg},
together with an in-house implementation of the {\tt OpenLoops} optimisation~\cite{Cascioli:2011va}.
In \mg, the computation is based on event reweighting, as described in
\cite{Frederix:2014hta,Maltoni:2014eza}. 
This functionality in \mg has since been automated as documented in
\cite{Mattelaer:2016gcx}. 
In practice, the HEFT is used to generate the events at NLO which are
then reweighted to introduce the full top quark mass dependence of the one- and two-loop amplitudes. 

\medskip

In order to study the phenomenological impact of the two different
matching schemes implemented in the \powhegbox{} and in \mg{}, we
compare the two results using the same parton shower. In both cases we
use {\tt Pythia~8.2} with the same settings to produce showered events
from both the \powhegbox{} and the {\tt MG5\_aMC@NLO} results at LHE
level. This means that the differences in the distributions produced
by \powheg{} and {\tt MG5\_aMC@NLO} respectively, will only be due to
the corresponding matching schemes.


%% file: results.tex
In this section we present phenomenological results and compare
predictions at different levels. We start presenting some consistency
checks at the fixed order level and at the Les Houches event level
(LHE), i.e. after the first hard emission is generated according to
the \powheg method. To assess the impact of the parton shower and
estimate its capacity to include approximate higher order effects,
in Section~\ref{sec:nnlo_effects} we compare results in the basic HEFT
approximation at NLO+PS with the NNLO predictions from
Reference~\cite{deFlorian:2016spz}. Finally, in
Section~\ref{sec:nlops_full} NLO and NLO+PS results in the full SM are
presented.

All the results we computed using the
PDF4LHC15{\tt\_}nlo{\tt\_}30{\tt\_}pdfas~\cite{Butterworth:2015oua,CT14,MMHT14,NNPDF}
parton distribution functions interfaced to our codes via
LHAPDF~\cite{Buckley:2014ana}, along with the corresponding value for
$\alpha_s$.  The masses of the Higgs boson and the top quark have been
set, as in the virtual amplitude, to $m_h=125$\,GeV, $m_t=173$\,GeV,
respectively, whereas their widths have been set to zero. As already
mentioned in the previous section, we consider on-shell Higgs bosons
and leave the analysis of more exclusive final states, stemming from
Higgs boson decays, to future studies.  Jets are clustered with the
anti-$\kt$ algorithm~\cite{Cacciari:2008gp} as implemented in the
\fastjet package~\cite{Cacciari:2005hq, Cacciari:2011ma}, with jet
radius $R=0.4$ and a transverse momentum greater than
$p_{T,min}^{\rm{jet}}=20~\GeV$.  The theoretical scale uncertainty is
estimated by varying the factorization scale $\mu_{F}$ and the
renormalization scale $\mu_{R}$. The scale variation bands are
obtained by computing the envelopes of a 7-point scale variation
around the central scale $\mu_0 =\mhh/2$, with
$\mu_{R,F}=c_{R,F}\,\mu_0$, where $c_R,c_F\in \{2,1,0.5\}$. The
extreme variations $(c_R,c_F)=(2,0.5)$ and $(c_R,c_F)=(0.5,2)$ have
been omitted.

We also have varied the PDFs using the $30$ error PDFs contained in the\\
PDF4LHC15{\tt\_}nlo{\tt\_}30{\tt\_}pdfas set and found that the
uncertainty due to PDF variations never exceeds 6\% and therefore is
well below the scale variation uncertainty.
The uncertainty bands
shown on our results originate from scale variations only.

We should point out that we switched off the hadronisation and the
multiple interactions in the parton shower.  A detailed
phenomenological study including various Higgs boson decay channels as
well as hadronisation effects will be left to a subsequent
publication.

\subsection{Comparison with previous NLO results}

A very strong consistency check of the new implementation, which
allows to test at the same time the real amplitudes and their
stability, the implementation of the grid for the virtual two-loop
amplitude and  all the various parts of the code
relevant for the NLO fixed order computation is a comparison with the
previous NLO results computed in
Refs.~\cite{Borowka:2016ehy,Borowka:2016ypz}.
A comparison at the level of individual phase space points shows that the
grid interpolation slightly increases the numerical uncertainty associated to
the virtual amplitude results, which were calculated with percent level
precision in the previous publications. At the level of differential
distributions we found excellent agreement, not only
for the full NLO results, but also for the various other
approximations available and for uncertainties related to scales
variation.

\begin{figure}[htb]
\centering
\begin{subfigure}{0.49\textwidth}
\includegraphics[width=\textwidth]{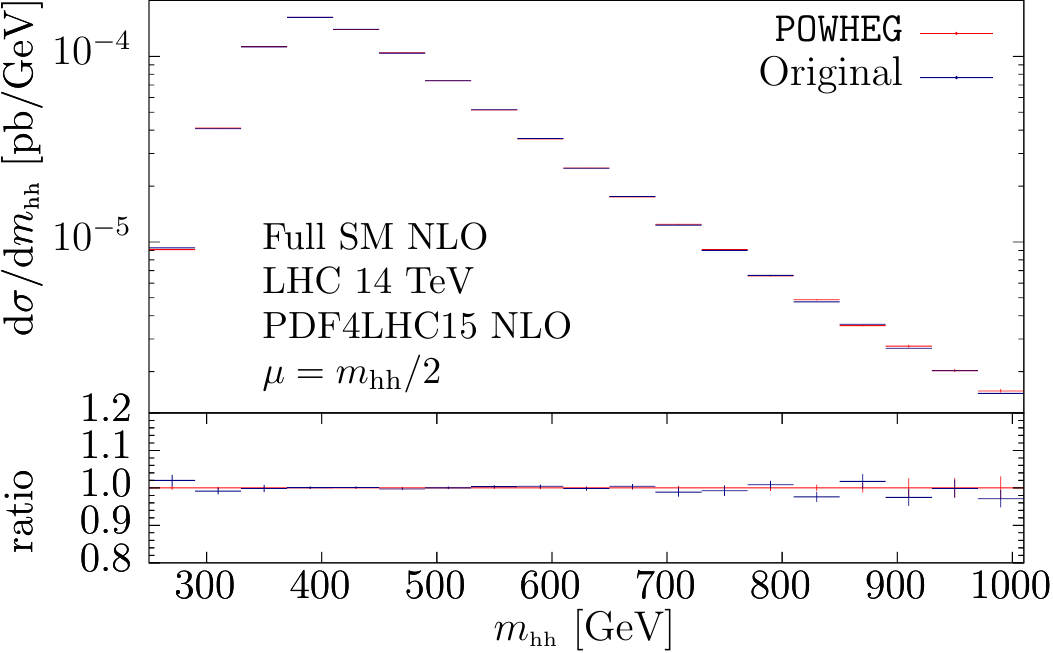}
\end{subfigure}
\begin{subfigure}{0.49\textwidth}
\includegraphics[width=\textwidth]{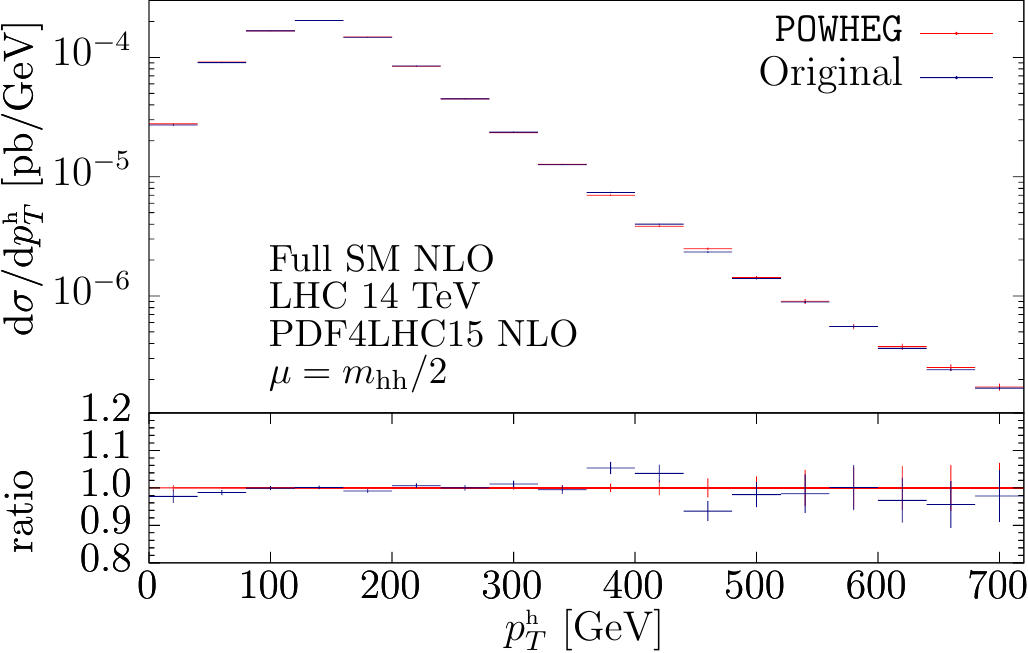}
\end{subfigure}
\caption{The $\mhh$ and $p_{T}^{\mathrm{h}}$ distributions calculated
  from the grid versus the full calculation. \label{fig:gridvsfull}}
\end{figure}

In Fig.~\ref{fig:gridvsfull} we show a comparison between
the NLO predictions obtained with the \powhegbox{} generator and the
original ones from Ref.~\cite{Borowka:2016ypz}.
In the \mg case, where only showered results are available, we made a similar validation plot for the 
$\mhh$ distribution, as it is insensitive to shower effects.

We should mention at this point that beyond
$\pth\sim 650~\GeV$, a systematic bias stemming from lack of statistics in the grid starts to develop.
As a consequence, the results obtained with the grid will be systematically below the ``true'' results. 
The difference is within the statistical uncertainty up to about $\pth\sim 750~\GeV$, and increases to 
about 20\% at $\pth\sim 1~\TeV$ (and $\mhh\sim 2~\TeV$).

\subsection{Comparisons at the level of Les Houches event files}

\begin{figure}
\centering
\begin{subfigure}{0.49\textwidth}
\includegraphics[width=\textwidth]{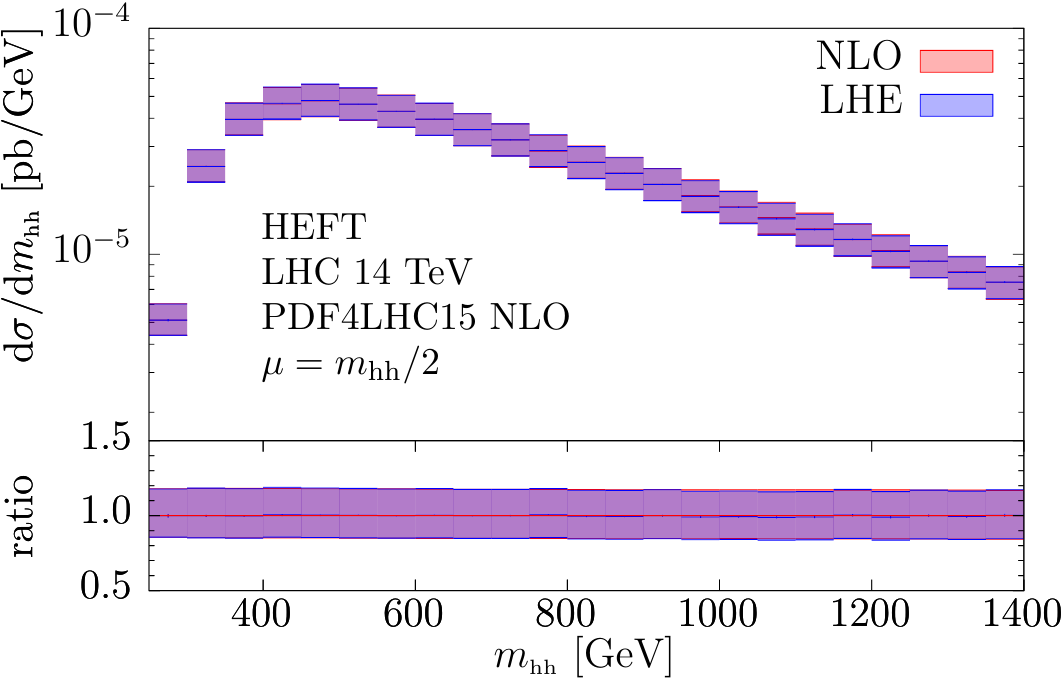}
\caption{\mhh in HEFT\label{fig:LHEmhh_heft}}
\end{subfigure}
\begin{subfigure}{0.49\textwidth}
\includegraphics[width=\textwidth]{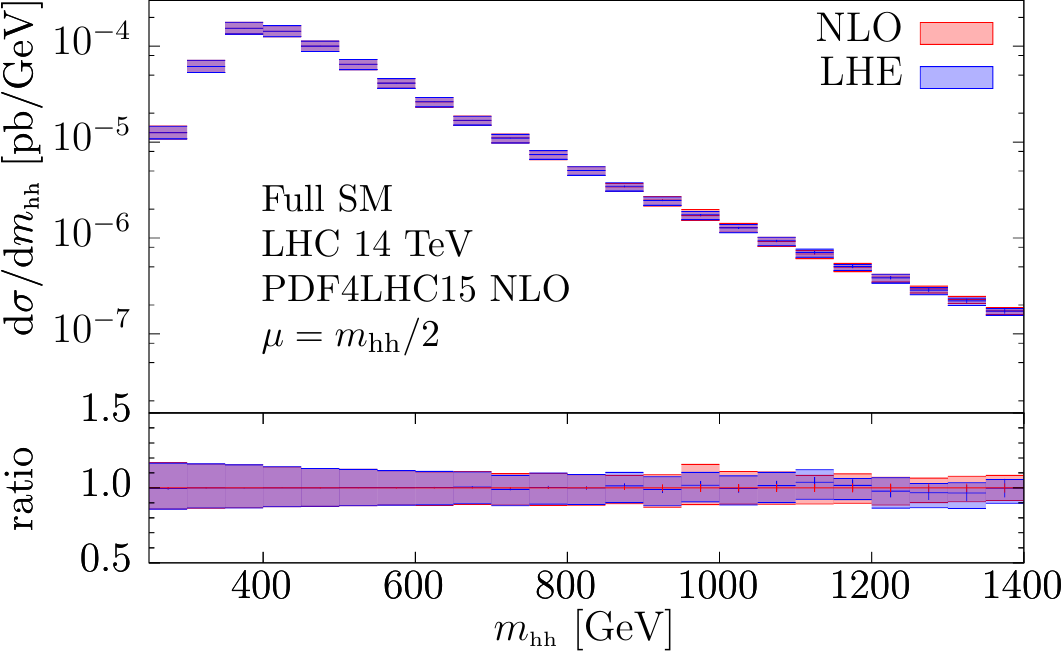}
\caption{\mhh in the full SM\label{fig:LHEmhh_ftap}}
\end{subfigure}
\caption{%
Higgs-pair invariant mass distributions \mhh in the HEFT
approximation and in the full SM at fixed NLO level compared to
LHE level, where in the latter the value $\hdamp=\infty$ has been used.\label{fig:mhh_NLOvsLHE}}
\end{figure}

\begin{figure}
\centering
\begin{subfigure}{0.49\textwidth}
\includegraphics[width=\textwidth]{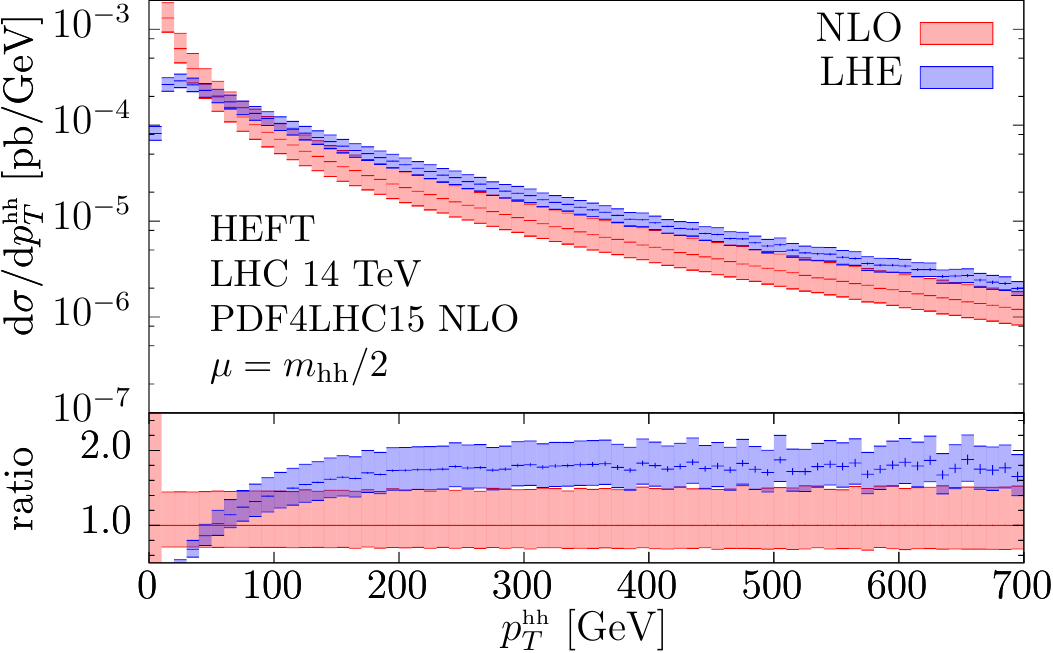}
\caption{\pthh in HEFT\label{subfig:LHEpthh_heft}}
\end{subfigure}
\begin{subfigure}{0.49\textwidth}
\includegraphics[width=\textwidth]{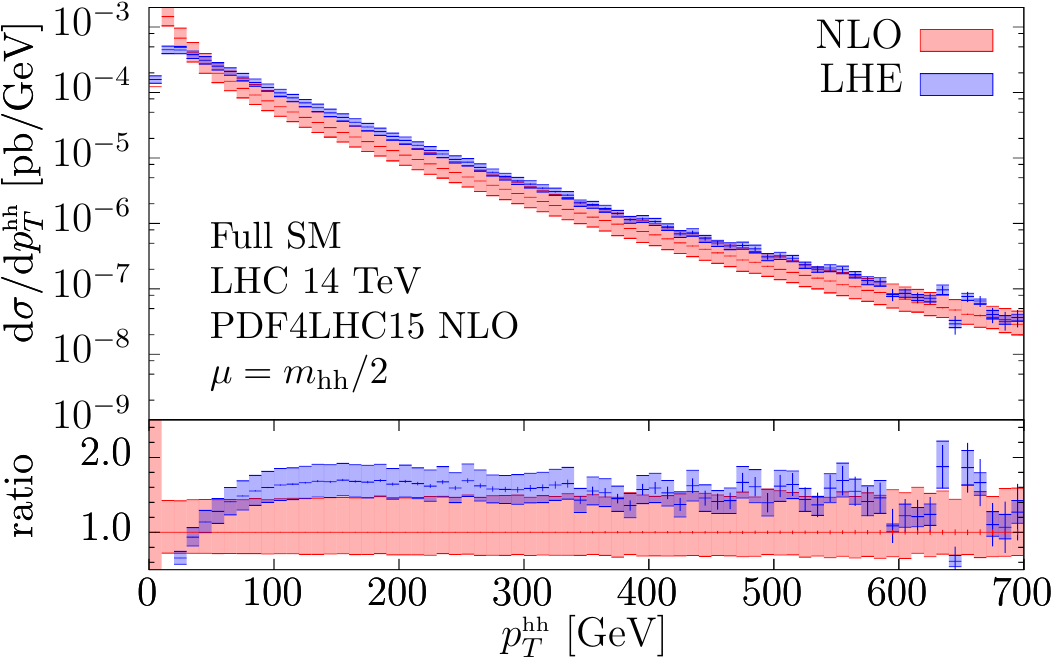}
\caption{\pthh in the full SM\label{subfig:LHEpthh_full}}
\end{subfigure}
\caption{%
  Higgs-pair transverse momentum distributions \pthh in the HEFT
  approximation and in the full SM at fixed NLO level compared to LHE
  level (with $\hdamp=\infty$). \label{fig:pthh_NLOvsLHE}}
\end{figure}

\begin{figure}
\centering
\begin{subfigure}{0.49\textwidth}
\includegraphics[width=\textwidth]{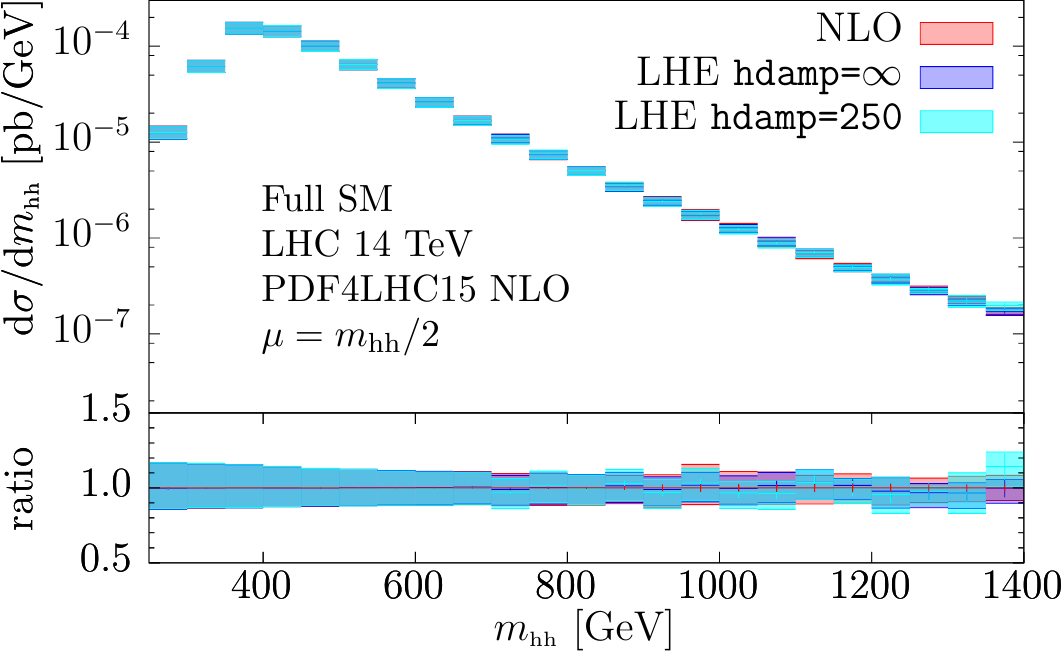}
\caption{\mhh in the full SM\label{subfig:LHEmhh_hdamp_full}}
\end{subfigure}
\begin{subfigure}{0.49\textwidth}
\includegraphics[width=\textwidth]{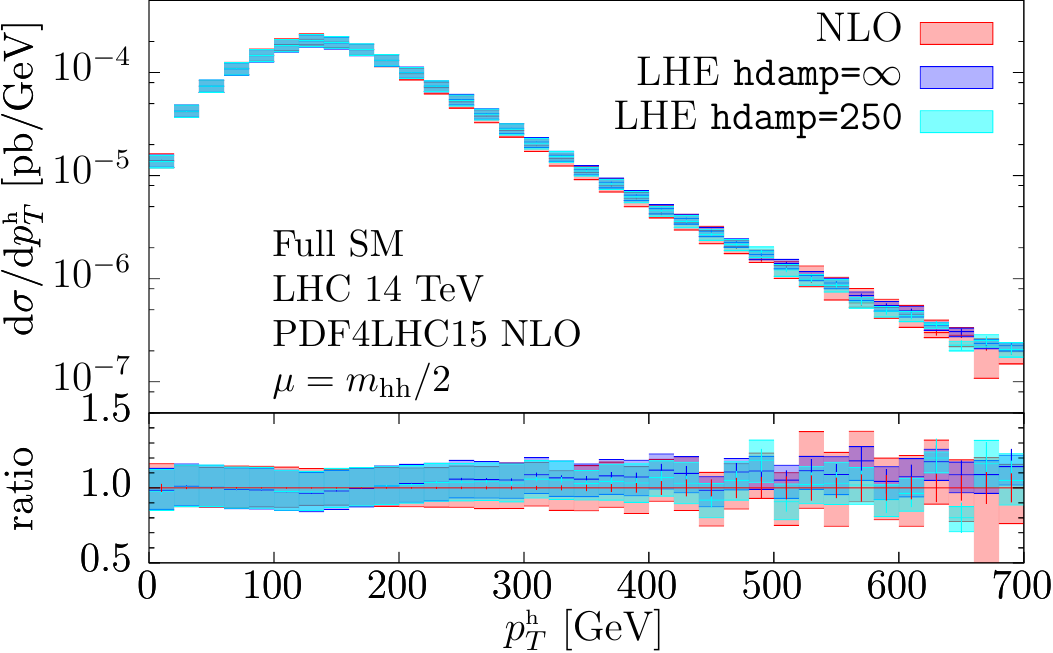}
\caption{$\pth$ in the full SM\label{subfig:LHEpth1_hdamp_full}}
\end{subfigure}
\caption{%
Comparison of the \powheg predictions with $\hdamp=\infty$ at LHE
level with predictions in which we set $\hdamp=250$, for the Higgs-pair
invariant mass $\mhh$ (left) and the transverse momentum of any
(randomly chosen) Higgs boson. The predictions are computed with full
top-quark mass dependence. \label{fig:mhh_hdamp_NLOvsLHE}}
\end{figure}

Before presenting results for NLO predictions matched to the parton
shower, we show comparisons of NLO curves with results at the Les
Houches event (LHE) level, after the first hard emission is weighted
with the Sudakov factor according to the \powheg method. Even though the
LHE level predictions still need to be showered, such a comparison allows to test
the implementation and, once the results are fully showered, to
disentangle the impact of the shower from the one due to the \powheg
exponentiation. For observables which are inclusive in the
extra radiation, the fixed order NLO and LHE level predictions should
be in perfect agreement. We show the level of agreement between the
NLO and LHE curves for the Higgs-boson pair invariant mass $m_{hh}$ in
Fig.~\ref{fig:mhh_NLOvsLHE}, where a comparison is shown for
predictions in the HEFT approximation and the full SM.
For observables which are directly sensitive to soft gluon radiation,
like the $\pthh$ distribution in the limit $\pthh\to 0$, one instead
expects to observe Sudakov suppression in the soft region. This can be
seen in Fig.~\ref{fig:pthh_NLOvsLHE}, where we can clearly see the
suppression in the region where the fixed order NLO results become
unreliable.
We also note that the LHE predictions are enhanced in the high
transverse momentum region compared to the NLO curve. This is due to
subleading contributions in the exponential, which in the case of
large radiative corrections can become sizable, in particular for
observables like \pthh, where NLO is the first non-trivial order to
describe the distribution. Analogous effects have already been
observed in several other similar processes with large
K-factors~\cite{Alioli:2008tz,Alioli:2009je,Alioli:2016xab,Hespel:2015zea}, and we
refer the interested reader to
Refs.~\cite{Alioli:2008tz,Alioli:2009je} for more details. We have
explored the possibility to limit the amount of hard radiation which
is exponentiated by changing the \hdamp parameter in \powheg. We
recall that this allows to divide the contributions of the real
radiation $R$ which are exponentiated in the Sudakov factor into a
singular part $R_{\mathrm{sing}}$ and a regular part
$R_{\mathrm{reg}}$, as follows:
\begin{align}
R_{\mathrm{sing}}=\,&R \times F\,,\\
R_{\mathrm{reg}}=\,&R \times (1-F)\,,
\end{align}
where the transition function $F$ is chosen to be
\begin{align}
F=\frac{h^{2}}{(\pthh)^2+h^{2}}\,.
\end{align}
\begin{figure}
\centering
\begin{subfigure}{0.49\textwidth}
\includegraphics[width=\textwidth]{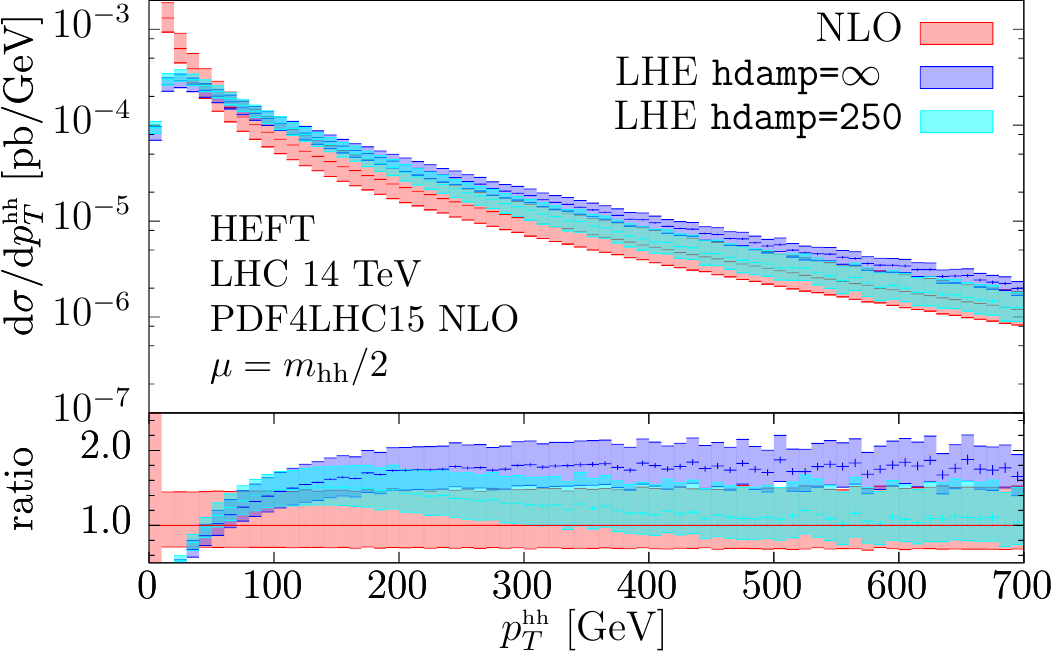}
\caption{$\pthh$ in basic HEFT\label{subfig:LHEpthh_hdamp_heft}}
\end{subfigure}
\begin{subfigure}{0.49\textwidth}
\includegraphics[width=\textwidth]{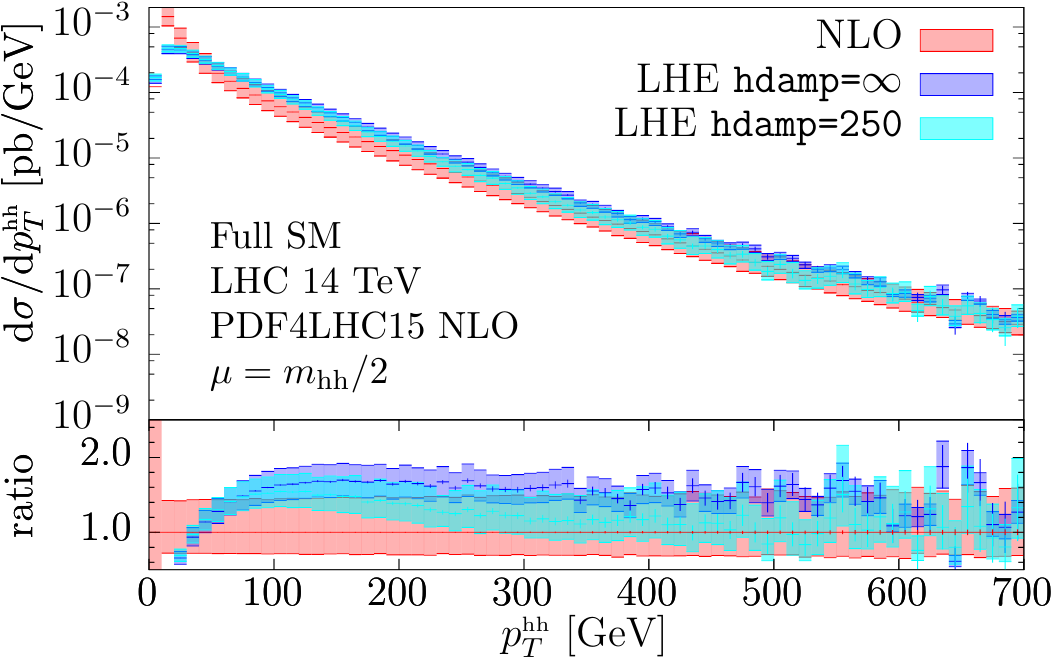}
\caption{$\pthh$ in the full SM\label{subfig:LHEpthh_hdamp_full}}
\end{subfigure}
\caption{%
  Comparison of the \powheg predictions with $\hdamp=\infty$ at
  LHE level with predictions in which we set $\hdamp=250$; for HEFT
  (left) and with full top-quark mass dependence
  (right). \label{fig:pthh_hdamp_NLOvsLHE}}
\end{figure}
In Fig.~\ref{fig:pthh_hdamp_NLOvsLHE} we compare the default \powheg
setting, $h=\hdamp=\infty$, with predictions where we use
$\hdamp=250~\GeV$. The left plot shows predictions in the HEFT, whereas on
the right we show results in the full SM. We observe that in both
cases above $500~\GeV$ the LHE curve with $\hdamp=250~\GeV$ reproduces the NLO results as
expected. It is interesting to study how this additional source of
theoretical uncertainty is affecting other observables, especially
those for which our predictions are NLO accurate. To understand this
better, in Fig.~\ref{fig:mhh_hdamp_NLOvsLHE} we show a similar
comparison for $\mhh$ (left) and the transverse momentum of
a (randomly chosen) Higgs boson $\pth$ (right), with full top quark mass
dependence. The $\mhh$ observable is completely insensitive to
additional radiation, and for this reason it is unaffected by a
modification of the \hdamp factor. This is not true for $\pth$, which
is sensitive to the recoil against additional jet activity. For this
reason we observe deviations between the NLO predictions and the
LHE-level curves, the latter  becoming slightly larger for harder transverse momenta. The
predictions for $\hdamp=250$ are in general closer to the NLO ones
over the whole kinematical range of $\pth$. We stress however that,
contrary to $\pthh$, where the differences between the predictions for
$\hdamp=\infty$ and the one for $\hdamp=250~\GeV$ reach 80\% above 500~\GeV, for
$\pth$ the differences are at the 10-15\% level, i.e. well within the
scale uncertainties.

Since the uncertainty related to the value for \hdamp is very tightly
related to the \powheg way of matching NLO to the parton-shower, it is
important to compare these predictions with other matching schemes. We
will comment more on this aspect in Section~\ref{sec:nlops_full}, where
we compare NLO+PS predictions obtained with
\powheg and \mg.

\subsection{Discussion of NNLO effects}\label{sec:nnlo_effects}

We mentioned in the previous section that the enhancement in the tail of
$\pthh$ in the prediction at the LHE-level is due to subleading
contributions in the Sudakov factor, which are intrinsically taken
into account in the matching {\`a} la \powheg. As already pointed out
and discussed in~\cite{Alioli:2008tz}, it is therefore interesting to
compare NLO+PS predictions obtained with \powheg to the full NNLO
predictions, if they are available. This is indeed the case if we
restrict ourselves to results in the basic HEFT, for which
differential NNLO results were computed in
Ref.~\cite{deFlorian:2016uhr}\footnote{We are grateful to Javier
Mazzitelli for providing us the NNLO predictions shown in the
comparisons of this section.}.
\begin{figure}
\centering
\begin{subfigure}{0.49\textwidth}
\includegraphics[width=\textwidth]{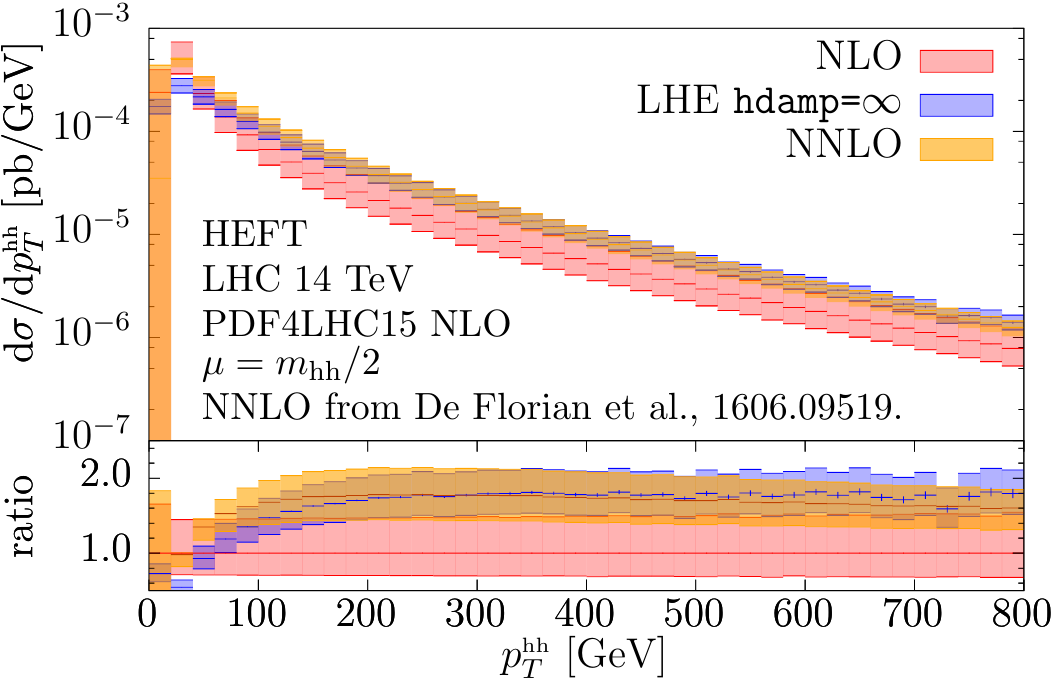}
\caption{$\pthh$ in basic HEFT, default value for $\hdamp$.\label{subfig:LHEpthh_nnlo_heft}}
\end{subfigure}
\begin{subfigure}{0.49\textwidth}
\includegraphics[width=\textwidth]{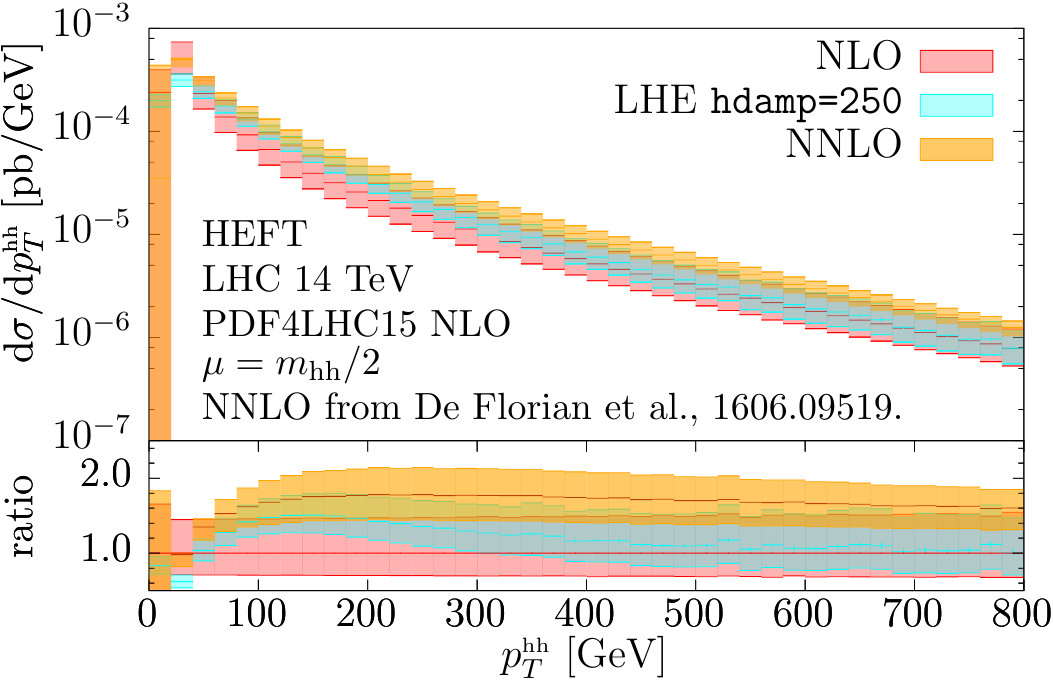}
\caption{$\pthh$ in basic HEFT, $\hdamp=250$.\label{subfig:LHEpthh_nnlo_hdamp_heft}}
\end{subfigure}
\caption{%
  Comparison of the NNLO results from Ref.~\cite{deFlorian:2016uhr}
  with default \powheg predictions ($\hdamp=\infty$) at LHE level
  (left) and predictions in which we set $\hdamp=250$ (right) for the
  Higgs-pair transverse momentum
  $\pthh$. \label{fig:pthh_nnlo_NLOvsLHE}}
\end{figure}
The comparison is of course meaningful only for those observables
which are LO accurate in our NLO calculation and which therefore are
not sensitive to the additional two-loop virtual corrections included
in the HEFT NNLO predictions. In Fig.~\ref{fig:pthh_nnlo_NLOvsLHE} we
consider again the transverse momentum of the Higgs boson pair and
compare the NNLO results with two different LHE-level predictions
from \powheg. On the left we keep the default setting in which
$\hdamp=\infty$, on the right we set $\hdamp=250~\GeV$. In the former
plot we observe a good agreement of the LHE-level curve with
$\hdamp=\infty$ with the NNLO predictions in the transverse momentum
range between $200~\GeV$ and $400~\GeV$. While the LHE-level result
flattens out around $250~\GeV$, the NNLO result decreases slightly for
larger $\pthh$. The two theory uncertainty bands due to scale
variation however largely overlap. The plot on the right shows instead
that, by limiting the amount of real radiation in the Sudakov factor,
the LHE-level prediction falls onto the NLO result at high $p_T$, and
therefore cannot reproduce the NNLO behaviour.

\begin{figure}
\centering
\begin{subfigure}{0.49\textwidth}
\includegraphics[width=\textwidth]{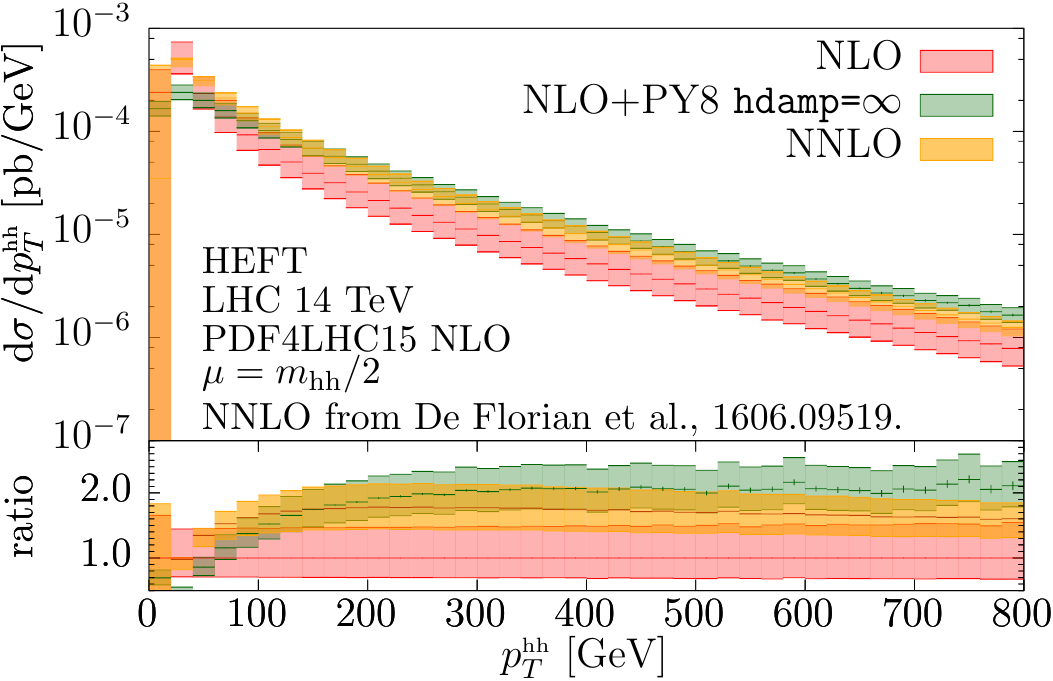}
\end{subfigure}
\begin{subfigure}{0.49\textwidth}
\includegraphics[width=\textwidth]{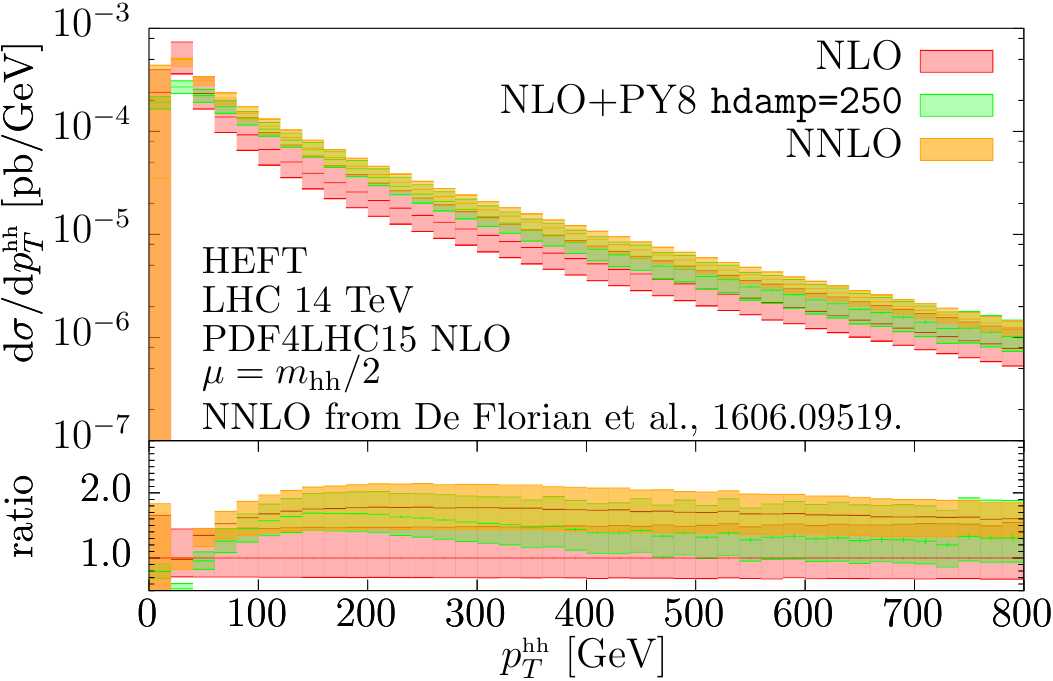}
\end{subfigure}
\caption{%
  Comparison of the NNLO results from Ref.~\cite{deFlorian:2016uhr}
  with default \powheg predictions ($\hdamp=\infty$) at NLO+PS level
  (left) and predictions in which we set $\hdamp=250$ (right) for the
  Higgs-pair transverse momentum $\pthh$. \pythia was
  used to shower the events. \label{fig:pthh_nnlo_NLOvsSWR}}
\end{figure}

As a further step, we can assess the impact of the parton shower, by
analyzing the same observable with NLO+PS predictions showered
with \pythia. Figure~\ref{fig:pthh_nnlo_NLOvsSWR} shows an analogous
comparison, where the NLO curves with and without shower are plotted
against the NNLO predictions. We observe that the shower has a large
effect on the tail of the $\pthh$ distribution, such that the NNLO
curve lies between the NLO+PS and the NLO fixed order curve for $\hdamp=\infty$.
On the other hand, for $\hdamp=250\,\GeV$, the NLO+PS result by construction is closer to the NLO fixed order result.
We should point out however that these considerations within the basic
HEFT approximation may not carry over analogously to the full calculation (where
NNLO predictions are not available), because it is well known that
the HEFT approximation does not have the correct scaling behaviour at
large transverse momenta.

\subsection{NLO plus parton shower matched results}\label{sec:nlops_full}

\begin{figure}
\centering
\begin{subfigure}{0.49\textwidth}
\includegraphics[width=\textwidth]{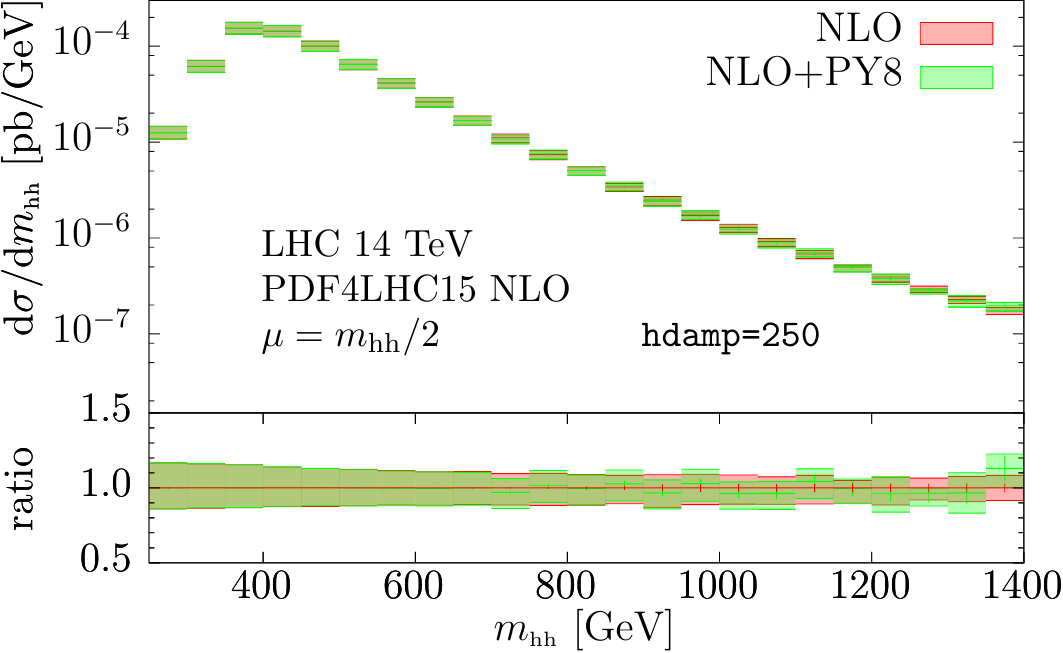}
\end{subfigure}
\begin{subfigure}{0.49\textwidth}
\includegraphics[width=\textwidth]{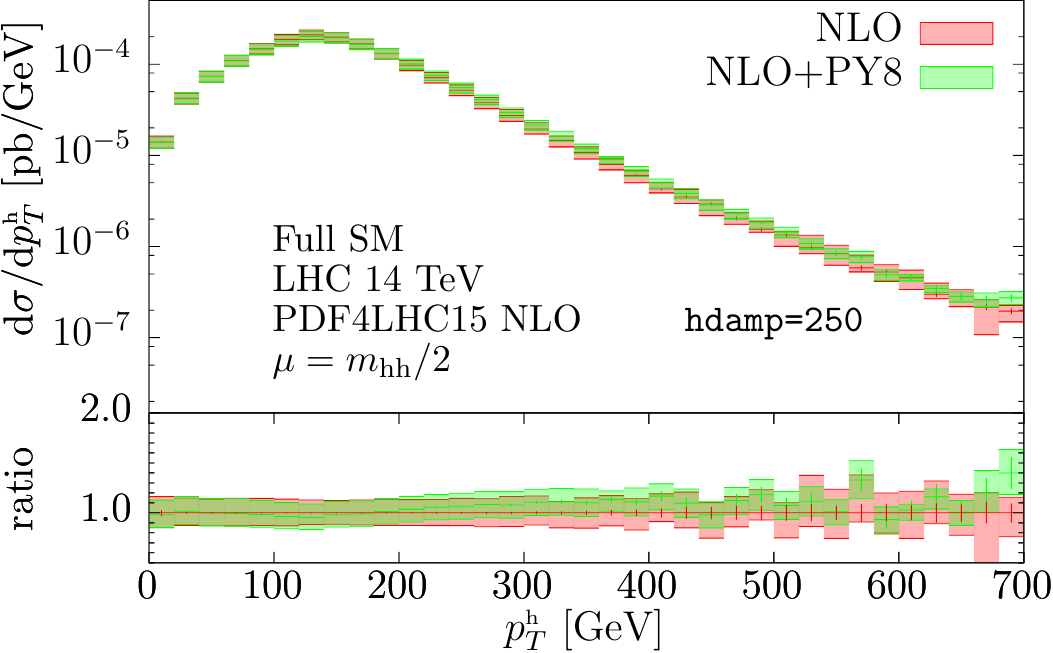}
\end{subfigure}
\caption{Higgs boson pair invariant mass distribution $m_{hh}$ and
  transverse momentum distribution of a (randomly chosen) Higgs boson at
  $\sqrt{s}=14$\,TeV, comparing the fixed order result with showered
  results from the \powhegbox{}.\label{fig:mhh}}
\end{figure}

\begin{figure}
\centering
\begin{subfigure}{0.49\textwidth}
\includegraphics[width=\textwidth]{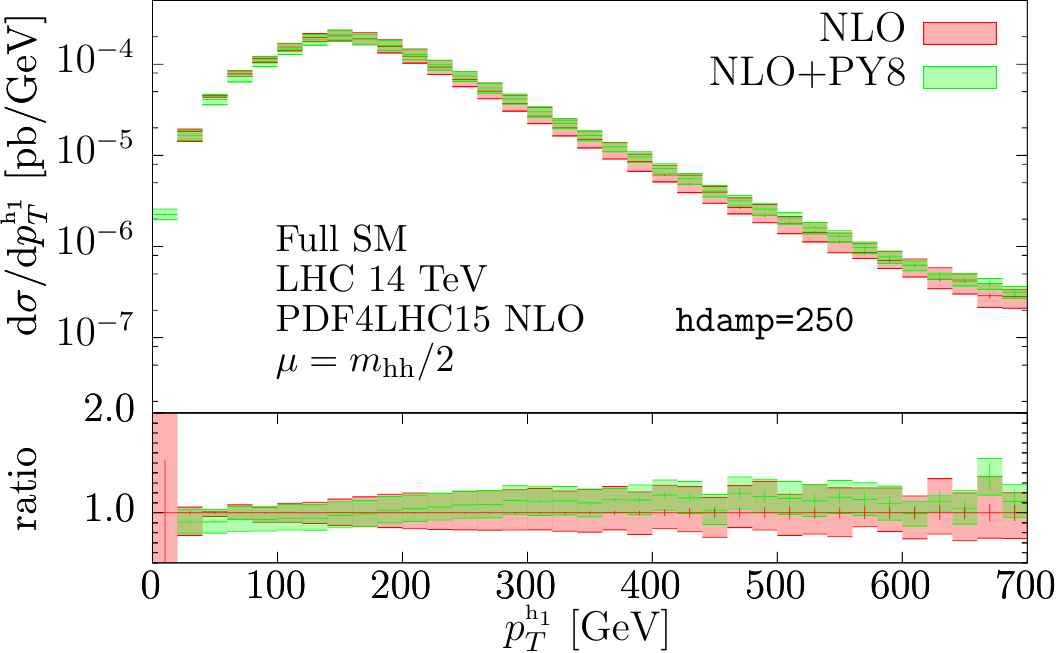}
\end{subfigure}
\begin{subfigure}{0.49\textwidth}
\includegraphics[width=\textwidth]{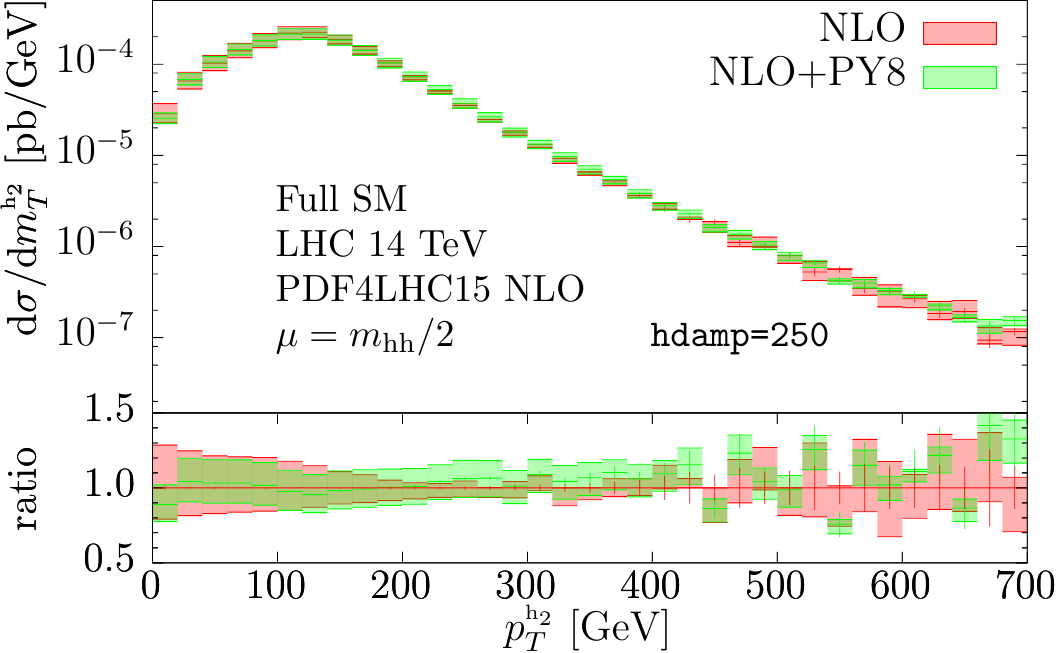}
\end{subfigure}
\caption{%
  Transverse momentum distribution of the leading ($\pthl$) and
  subleading ($\pths$) Higgs boson, comparing fixed order and showered
  results. The first bin in $\pthl$ in the fixed order NLO calculation is negative and therefore does not appear in the upper part of the plot.\label{fig:pthardsoft}}
\end{figure}

We now compare fixed order NLO results to our default \powheg results,
where we use \texttt{\hdamp=250} and  the \pythia shower. In
Fig.~\ref{fig:mhh} we show the Higgs boson pair invariant mass
distribution and the transverse momentum distribution of a randomly
chosen Higgs boson for both the fixed order and the showered
calculation.  As is to be expected, the invariant mass distribution is
rather insensitive to the parton shower.
In Fig.~\ref{fig:pthardsoft}  the $p_T$-distributions
of the harder and softer Higgs boson are shown.

\begin{figure}
\begin{center}
\includegraphics[width=0.55\textwidth]{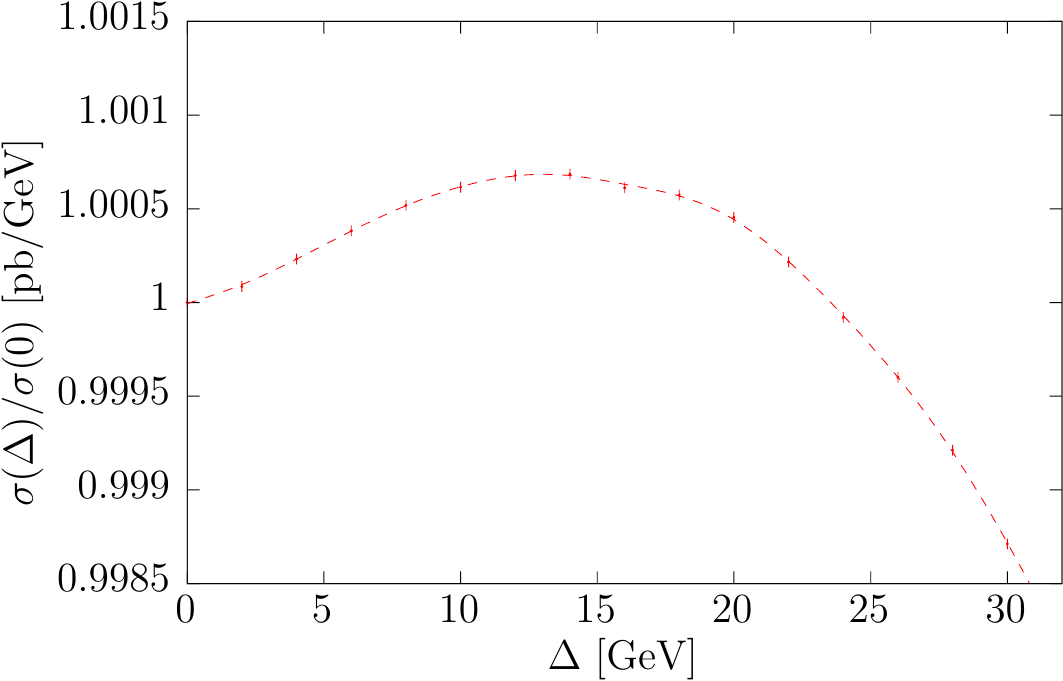}
\caption{Total cross section as a function of the difference $\Delta$
  between the $p_{T,\mathrm{min}}$ cut placed on the harder and the
  softer Higgs boson transverse momenta.\label{fig:deltah1h2}}
\end{center}
\end{figure}

We should mention at this point that the distributions of the
``harder'' ($\pthl$) and ``softer'' ($\pths$) Higgs boson,
calculated at fixed (NLO) order,  are somewhat
infrared sensitive if no cuts are placed on the Higgs boson transverse
momenta.
The reason is that, if the transverse momenta $\pthl$ and $\pths$ are
very close to each other, the available phase space for the extra
radiation in the real corrections is severely restricted, leading to
large logarithms which are not sufficiently balanced by the $2\to 2$
contributions. To illustrate this fact, we consider the
total cross section as a function of
$\Delta$, with the kinematic requirements $\pthl\ge \Delta, \pths \ge 0$.
The cross section shows an unphysical
behaviour as $\Delta\to 0$, see Fig.~\ref{fig:deltah1h2}:
the total cross section as a function of $\Delta$  peaks around $\Delta=14~\GeV$
and then decreases for smaller values of  $\Delta$,
even though the available phase space for $\pthl$ is larger.
This behaviour is an artifact of the fixed order calculation and
is the reason why ``symmetric cuts'' (i.e. the same $p_{T,min}$ values
for both final state particles in a $2\to 2$ calculation at NLO)
should be avoided. For a more detailed discussion of this point we
refer to Refs.~\cite{Catani:1997xc,Frixione:1997ks,Fontannaz:2001nq}.
Here we only note that this is the reason why, with ``symmetric'' cuts
$p_{T,min}^{\mathrm{h}_1} =p_{T,min}^{\mathrm{h}_2} =0$ and fine binning, the
first bin(s) of the $\pthl$ distribution are negative at fixed order,
while this behaviour is cured by the Sudakov factor, so it is absent
in the LHE level and showered results. 

\vspace*{3mm}

\begin{figure}
\centering
\begin{subfigure}{0.49\textwidth}
\includegraphics[width=\textwidth]{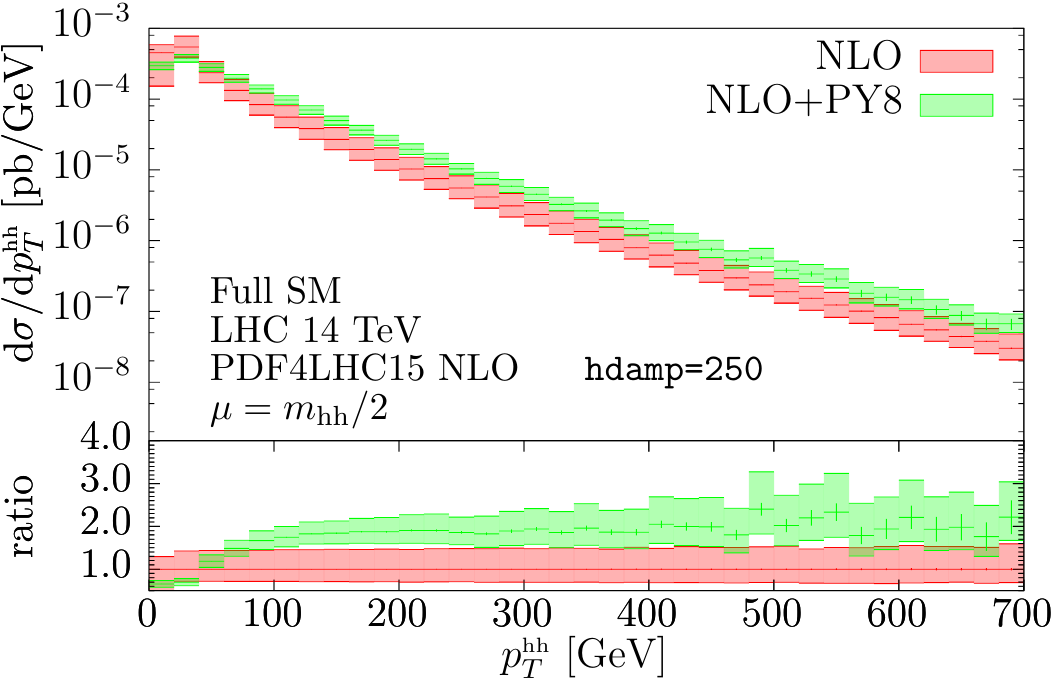}
\end{subfigure}
\begin{subfigure}{0.49\textwidth}
\includegraphics[width=\textwidth]{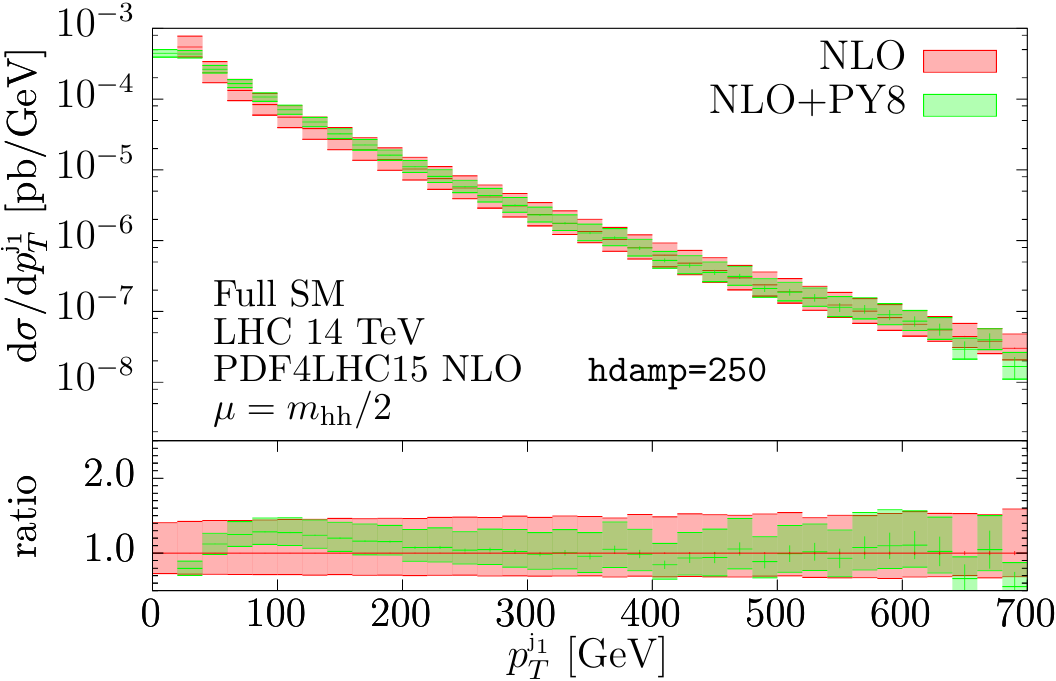}
\end{subfigure}
\caption{%
  Higgs boson pair transverse momentum distribution $\pthh$ (left) and
  leading jet transverse momentum distribution $\ptj$ (right),
  comparing fixed order and showered results.\label{fig:pthh}}
\end{figure}

Fig.~\ref{fig:pthh} displays the transverse momentum distributions of
the Higgs boson pair and
of the (leading) jet.  As discussed already in the context of
Fig~\ref{fig:pthh_hdamp_NLOvsLHE}, the $\pthh$ distribution diverges at fixed order for
$\pthh\to 0$, while the showered result is able to provide reliable
predictions in the low $\pthh$ region.
We notice that the scale variation band is reduced in the showered result compared to the fixed order calculation.
The scale uncertainties on the fixed order results are particularly large for these distributions as they are --
except for the first bin --
determined by the $2\to 3$ kinematics, which is described only at leading order accuracy by our calculation.

\begin{figure}
\centering
\begin{subfigure}{0.49\textwidth}
\includegraphics[width=\textwidth]{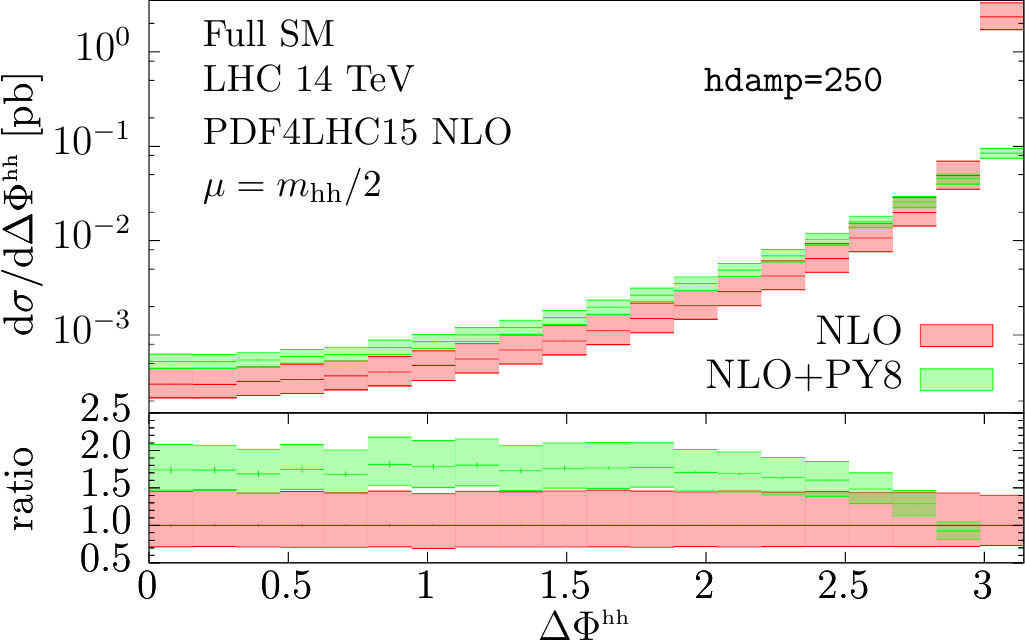}
\end{subfigure}
\begin{subfigure}{0.49\textwidth}
\includegraphics[width=\textwidth]{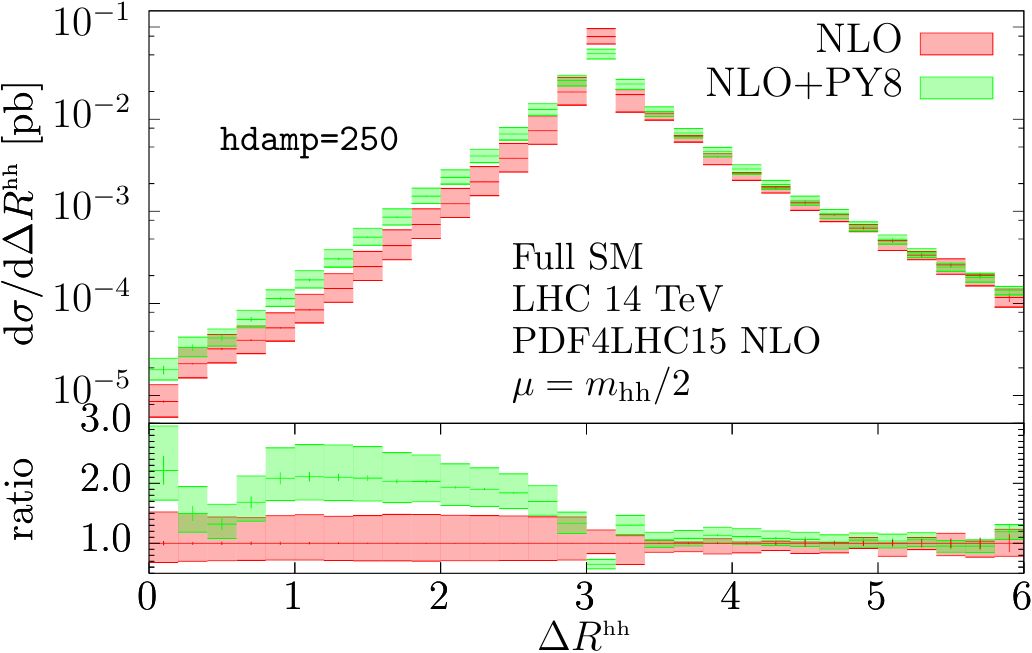}
\end{subfigure}
\caption{%
  Azimuthal angle separation $\dphihh$ (left) and radial separation
  $\drhh$ (right) of the two Higgs bosons, comparing fixed order and
  showered results.\label{fig:deltaPhihh}}
\end{figure}

In Fig.~\ref{fig:deltaPhihh} we show the difference in azimuthal
angle, $\Delta\Phi^{hh}$, and the  radial separation, $\drhh=\sqrt{(\eta_1-\eta_2)^2+(\Phi_1-\Phi_2)^2}$, of the two Higgs bosons.
We see that the unphysical behaviour for $\Delta\Phi^{hh}\to 0$ of the fixed order
result is cured by the Sudakov form factor, and again the scale uncertainties of the fixed order calculation
are relatively large because the tail of the distribution is predicted at the first non-trivial order.
In the  $\drhh$ distribution, we observe that the shower populates the region $\drhh <\pi$,
which at fixed order is given by the $2\to 3$  component only.

\begin{figure}
\centering
\begin{subfigure}{0.49\textwidth}
\includegraphics[width=\textwidth]{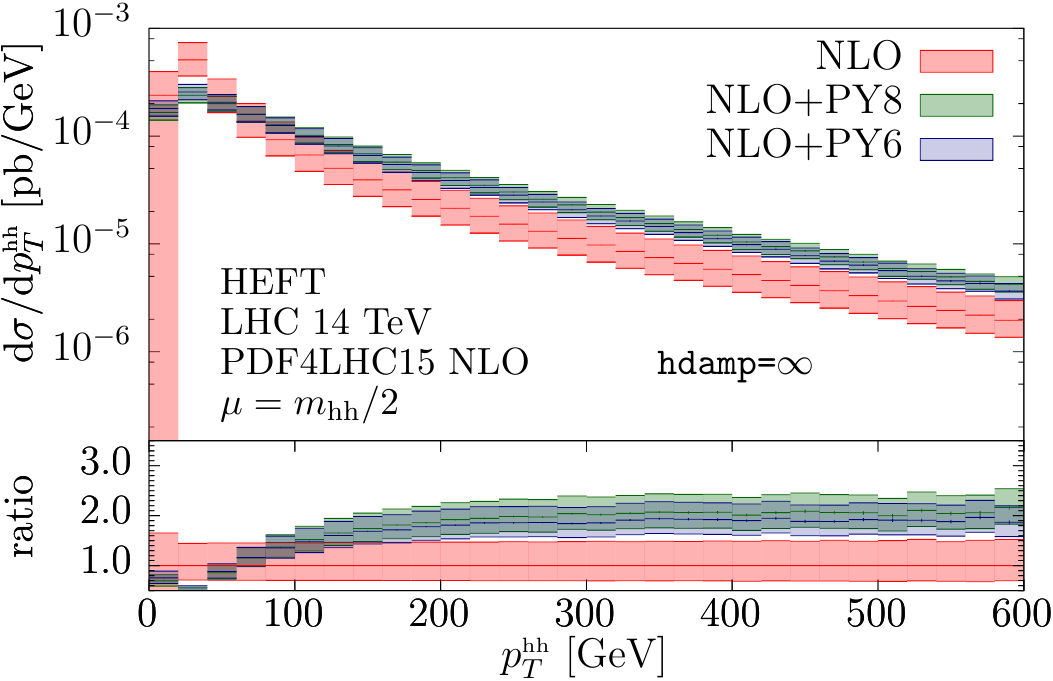}
\end{subfigure}
\begin{subfigure}{0.49\textwidth}
\includegraphics[width=\textwidth]{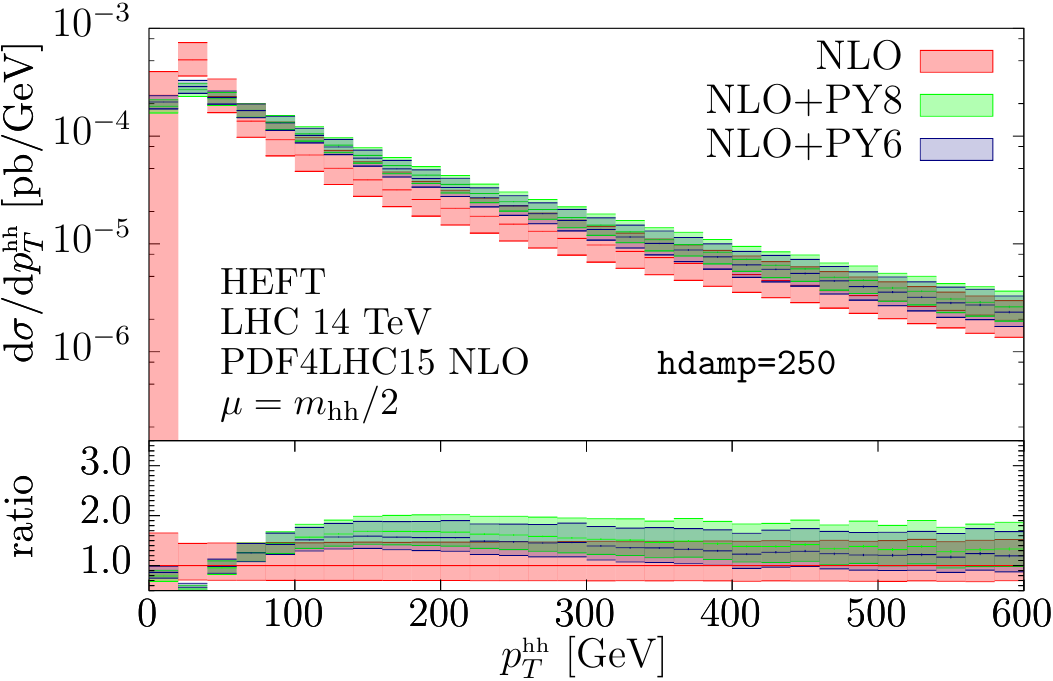}
\end{subfigure}
\\
\begin{subfigure}{0.49\textwidth}
\includegraphics[width=\textwidth]{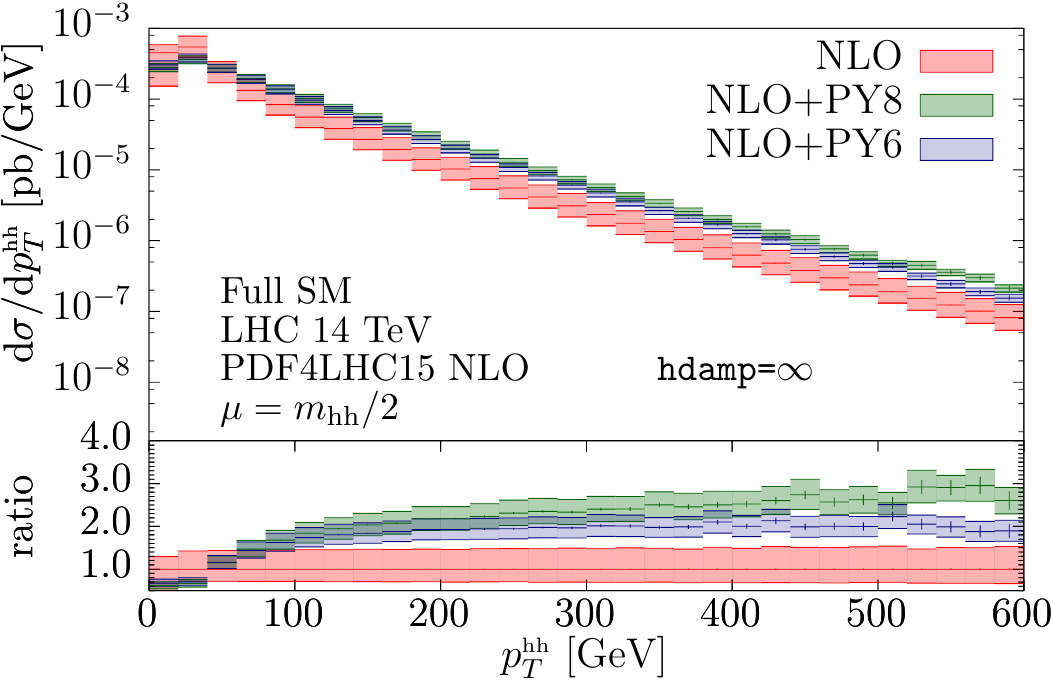}
\end{subfigure}
\begin{subfigure}{0.49\textwidth}
\includegraphics[width=\textwidth]{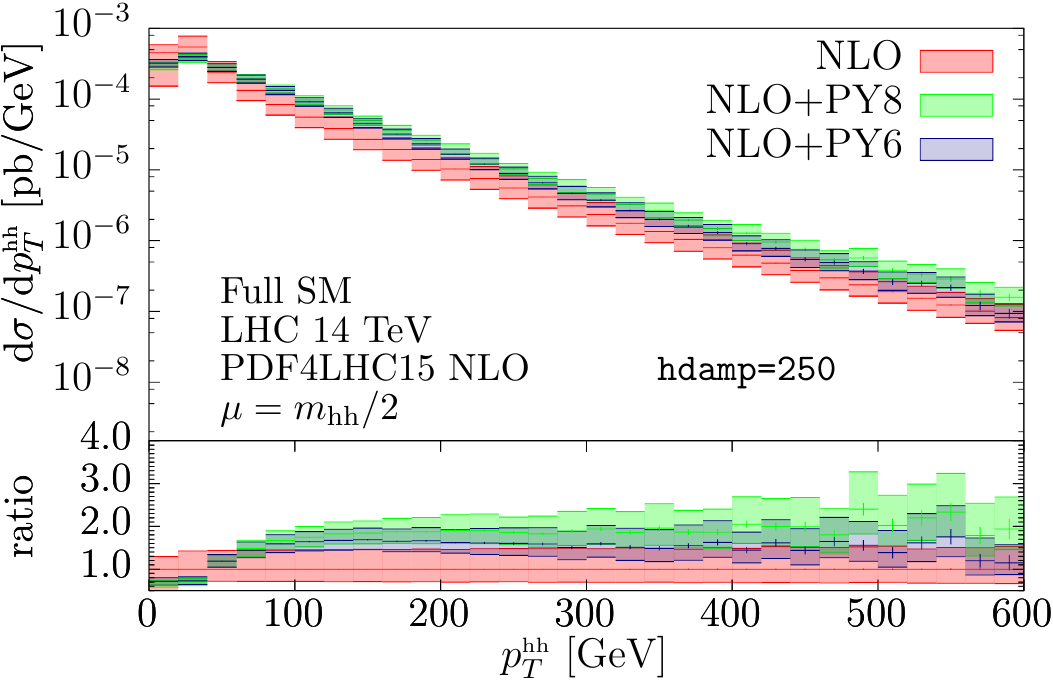}
\end{subfigure}
\caption{%
  Higgs boson pair transverse momentum distribution $\pthh$ (left column with \texttt{hdamp=$\infty$}, right column
  with \texttt{hdamp=250}) comparing the fixed order result with
  showered results from both \pythiaold and \pythia in the basic HEFT
  approximation (upper row) and in the full SM (lower
  row).\label{fig:pythia6}}
\end{figure}

Fig.~\ref{fig:pythia6} compares the predictions obtained with
the \pythiaold shower to the \pythia results both in the basic HEFT
approximation and in the full SM. It is instructive to make this
comparison for \texttt{hdamp=$\infty$} (left column) as well as
for \texttt{hdamp=250} (right column). In the basic HEFT approximation
the differences between \pythiaold and \pythia are small, and
setting \texttt{hdamp} to a finite value restores the agreement
between the NLO and the NLO+PS curves at large transverse
momentum. The latter is also true  in the full SM. However, 
in the full SM,  the difference between 
\pythiaold and \pythia is much larger, \pythia
showing a considerably harder spectrum in the  tail of the $\pthh$ distribution.


\begin{figure}
\begin{subfigure}{0.49\textwidth}
\includegraphics[width=\textwidth]{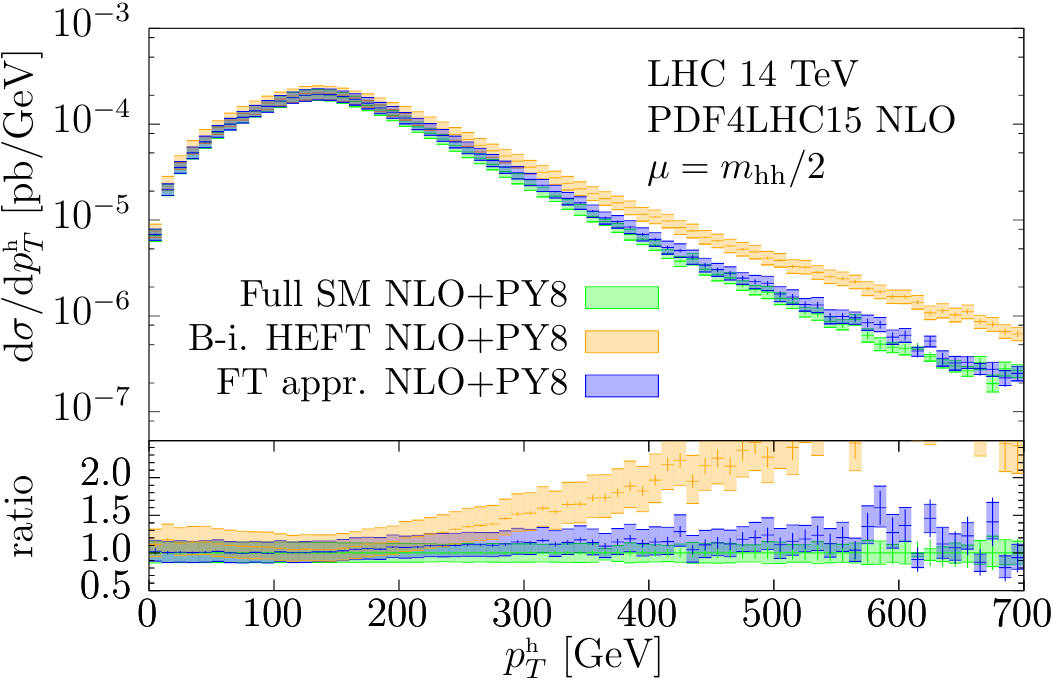}
\caption{Showered results ({\tt hdamp}=250).\label{subfig:approx_shower}}
\end{subfigure}
\begin{subfigure}{0.49\textwidth}
\includegraphics[width=\textwidth]{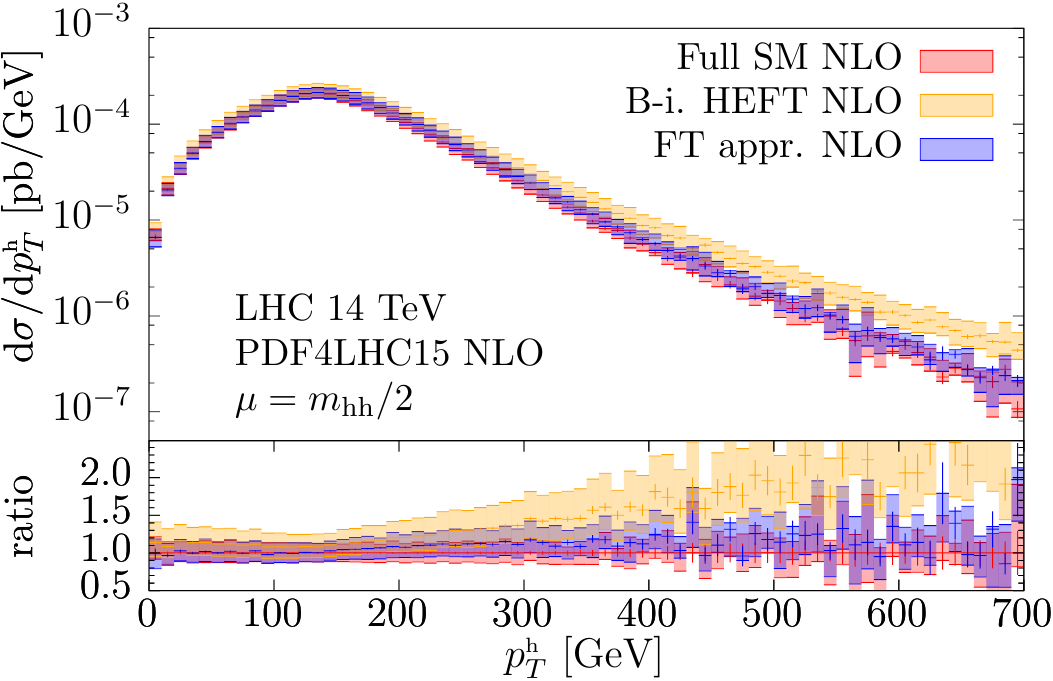}
\caption{Fixed order results.\label{subfig:approx_noshower}}
\end{subfigure}
\caption{$p_T^{h}$ distribution comparing (a) showered results based
  on matrix elements in various approximations (full, \ftapprox,
  Born-improved HEFT) with (b) fixed order
  results. \label{fig:pth_approx}}
\end{figure}

To conclude this section, in Fig.~\ref{subfig:approx_shower} we
compare the full results with calculations where the underlying matrix
elements are based on two approximations, either \ftapprox or
Born-improved HEFT.  All matrix elements are combined with the same
\pythia shower.  In order to assess the effect of the parton shower on
the various approximations, we also show the fixed order results in
Fig.~\ref{subfig:approx_noshower}.  The broad features of these
approximations remain unchanged after showering, however, as the
showered results have smaller scale uncertainties, the differences
between these approximations are actually enhanced if a parton shower
is attached.

\subsubsection{Comparison between \powheg and \mg}

In this section we compare the \powheg results with results from \mg,
the latter being based on the same grid in the invariants $\hat{s}$
and $\hat{t}$ for the virtual two-loop corrections as the \powheg
results, and based on the same \pythia shower.  Therefore the
differences between the results can be attributed to differences in
the matching scheme.

\begin{figure}
\centering
\begin{subfigure}{0.49\textwidth}
\includegraphics[width=\textwidth]{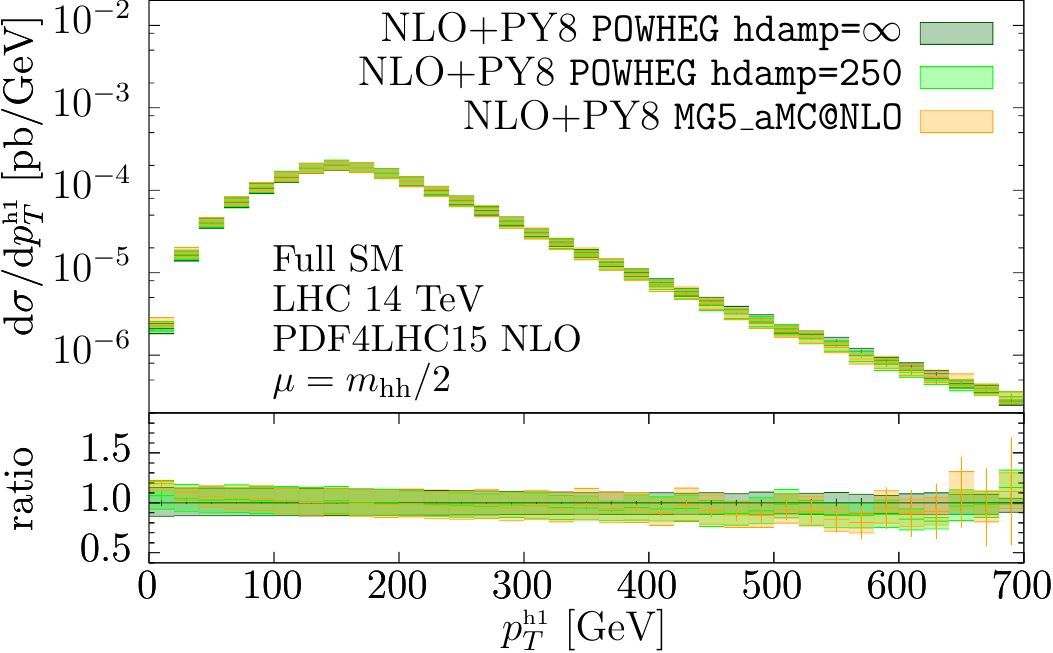}
\end{subfigure}
\begin{subfigure}{0.49\textwidth}
\includegraphics[width=\textwidth]{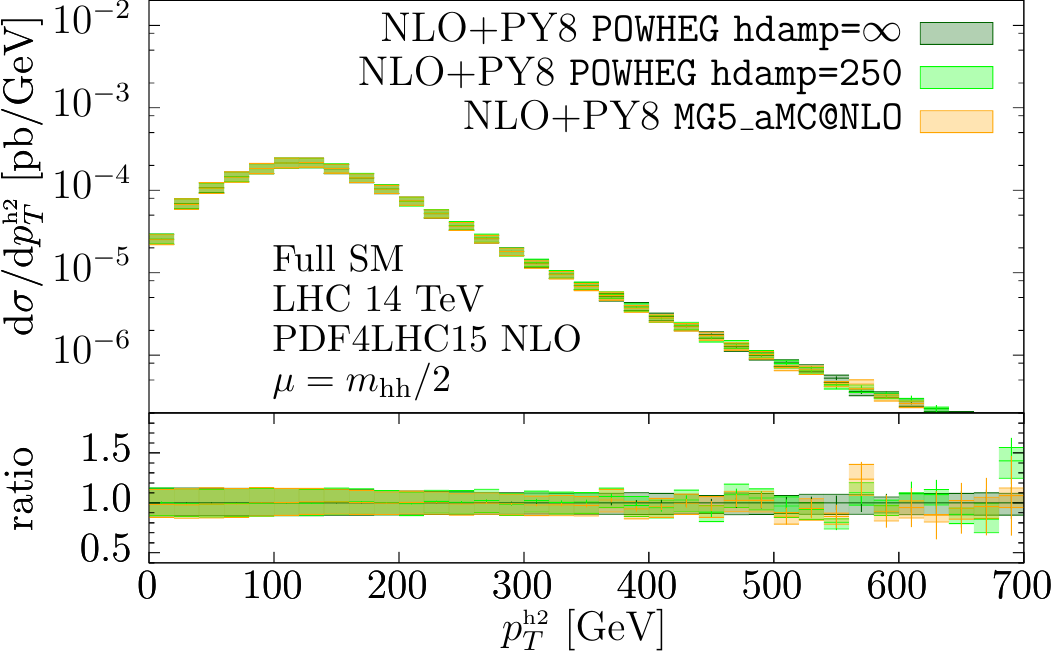}
\end{subfigure}
\caption{%
 Leading and subleading Higgs boson  transverse momentum distributions $\pthl$ and $\pths$, comparing
  showered results with \powheg and \mg.
 \label{fig:compMG5_pth1pth2}}
\end{figure}

\begin{figure}
\centering
\begin{subfigure}{0.49\textwidth}
\includegraphics[width=\textwidth]{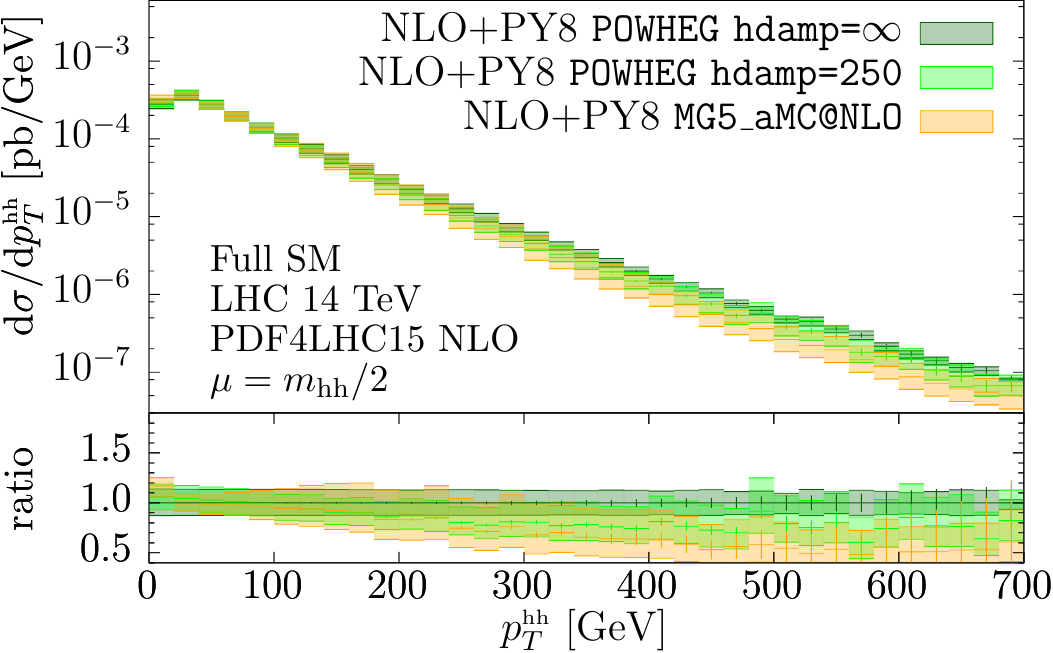}
\end{subfigure}
\begin{subfigure}{0.49\textwidth}
\includegraphics[width=\textwidth]{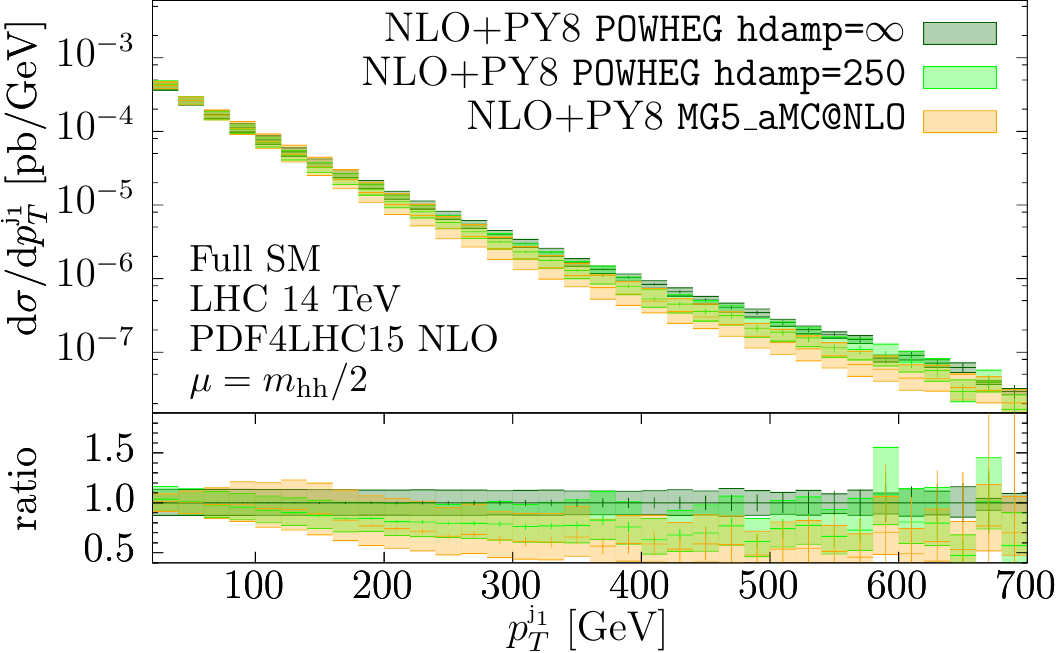}
\end{subfigure}
\caption{%
  Higgs boson pair transverse momentum distribution $\pthh$ (left) and
  $\ptj$ distribution (right), comparing
  showered results with \powheg and \mg.
  For the $\ptj$ distribution we used a cut of $p_{T,min}^{\rm{jet}}=20$\,GeV.
 \label{fig:compMG5_pthh}}
\end{figure}

\begin{figure}
\centering
\begin{subfigure}{0.49\textwidth}
\includegraphics[width=\textwidth]{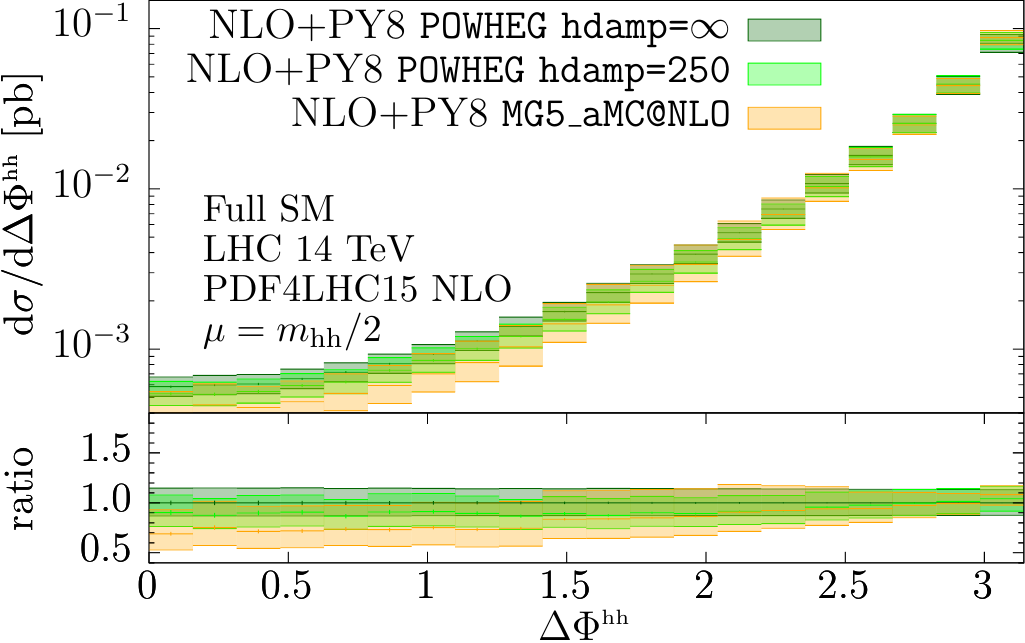}
\end{subfigure}
\begin{subfigure}{0.49\textwidth}
\includegraphics[width=\textwidth]{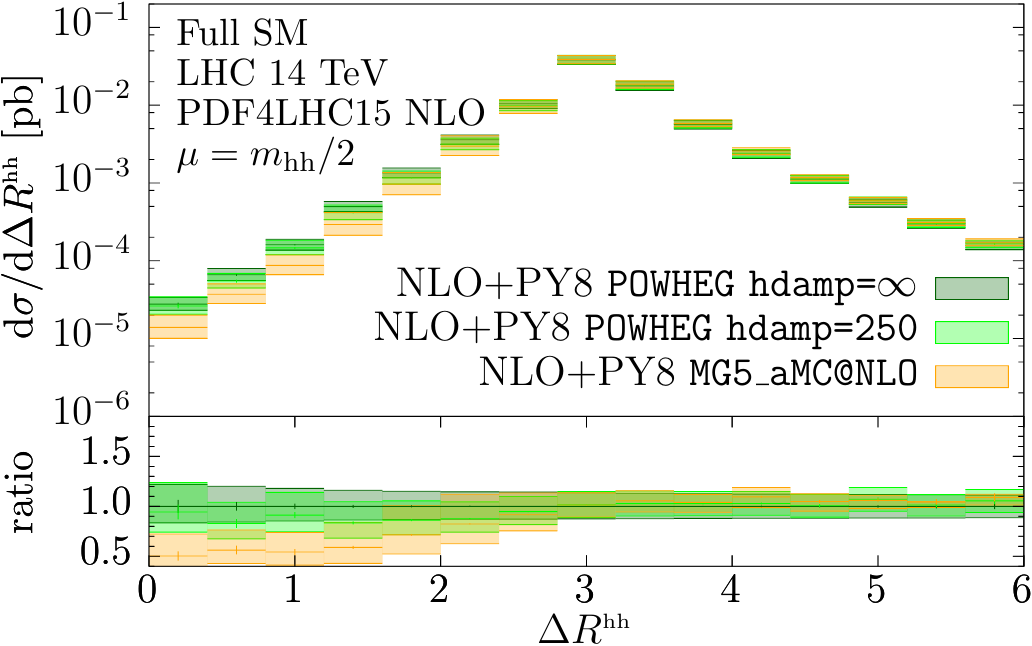}
\end{subfigure}
\caption{
 Azimuthal angle separation $\dphihh$ (left) and separation $\drhh$ (right).\label{fig:compMG5_pdphidr}}
\end{figure}

In Figs.~\ref{fig:compMG5_pth1pth2} to \ref{fig:compMG5_pdphidr} we
show \powheg results for two different values of {\tt hdamp} compared
to \mg results. While for the $\pthl$ and $\pths$ distributions the
differences are mostly small, they are, as to be expected, more
pronounced for the distributions where the shower populates kinematic
regions which are predicted at the first non-trivial order by the NLO
fixed order calculation.  Focusing on the comparison between the
\powheg curve with {\tt hdamp}$=250$, which is our default, we can
say that in general the two predictions agree well within the scale
and statistical uncertainties.  The small $\drhh$ region, on the right
of Fig.~\ref{fig:compMG5_pdphidr}, shows the largest differences,
which is not surprising as it is dominated by (multi-)jet events.  We
should also mention that the curve for {\tt hdamp}$=250$ in these
figures is close to the NLO curves by construction, as can be seen by
comparing to the fixed order results shown in the previous subsection.

\begin{figure}
\centering
\begin{subfigure}{0.49\textwidth}
\includegraphics[width=\textwidth]{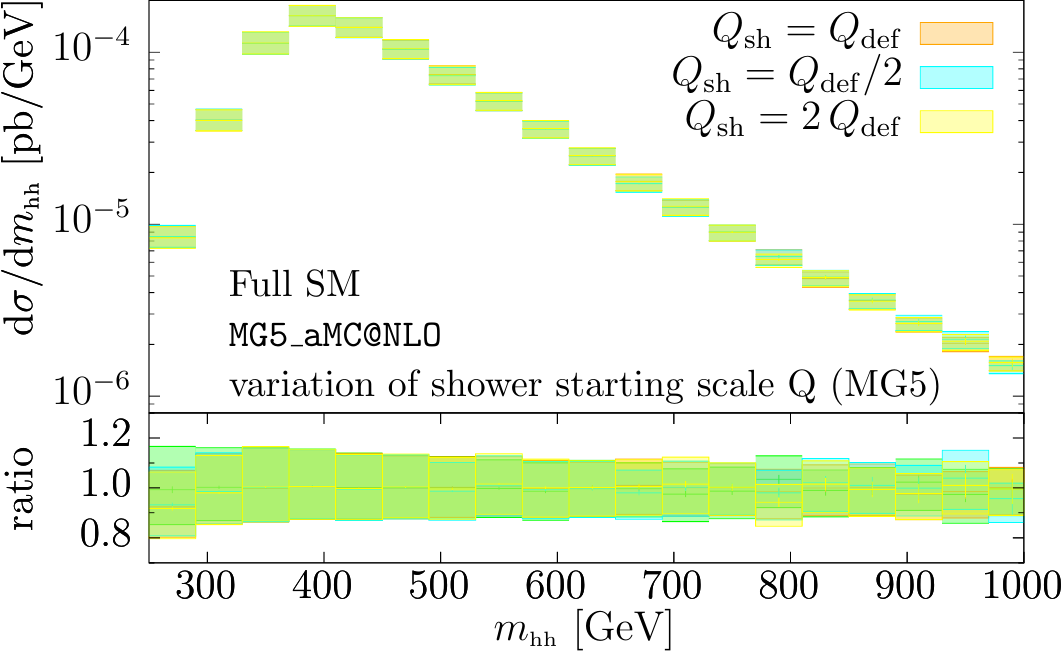}
\end{subfigure}
\begin{subfigure}{0.49\textwidth}
\includegraphics[width=\textwidth]{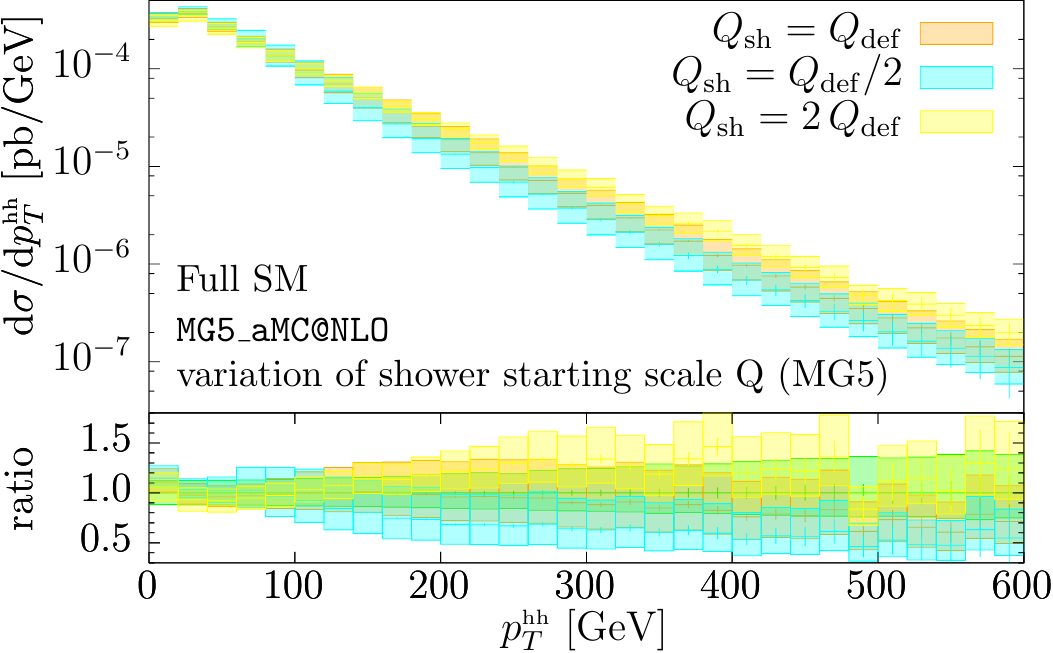}
\end{subfigure}
\caption{$\mhh$ and $\pthh$  distributions comparing  showered results
  based on the same matrix elements (NLO with full top quark mass
  dependence), varying the shower starting scale $Q_{\rm{sh}}$ in \mg{} by a factor of two up and down.
  The ratio plot is normalized to the \powheg result for {\tt hdamp=}250.
  The bands show the envelope of the variation of the renormalisation and factorisation scales.\label{fig:varyQ}}
\end{figure}

In Fig.~\ref{fig:varyQ} we vary the shower starting scale $Q_{sh}$ in
\mg  by a factor of two around the default
value.  
In the latest version of \mg (version 2.5.3 onwards), 
the shower starting scale is picked with some probability distribution
to be in the interval {\tt shower\_scale\_factor}
$\times \,[0.1\, H_T/2,H_T/2]$ with $H_T$ computed with Born kinematics, 
therefore to perform the scale variation we set the {\tt  shower\_scale\_factor} in the run card to 0.5, 1 and 2.

The $\mhh$ distribution can be considered as a control plot to demonstrate that, as expected,
this has no effect on the $\mhh$ distribution. In contrast, in the $\pthh$ distribution,
the differences due to variations of the matching scale start to exceed the scale uncertainties
towards larger $\pthh$ values.

\medskip

\begin{figure}
\centering
\begin{subfigure}{0.49\textwidth}
\includegraphics[width=\textwidth]{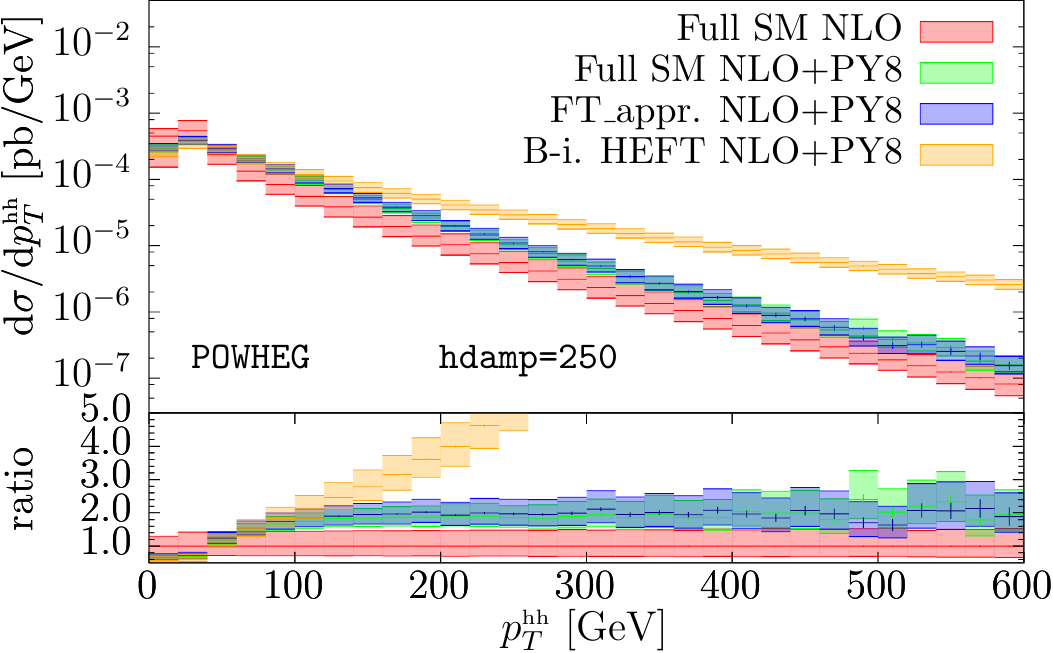}
\end{subfigure}
\begin{subfigure}{0.49\textwidth}
\includegraphics[width=\textwidth]{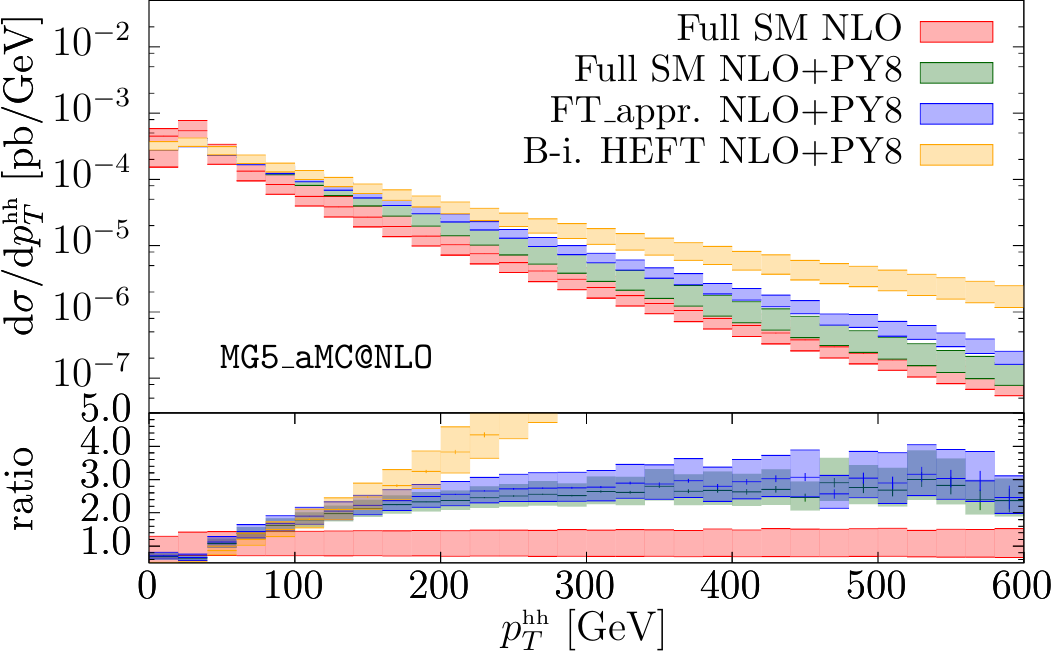}
\end{subfigure}
\caption{%
  Higgs boson pair transverse momentum distribution $\pthh$
  comparing fixed order and showered results. Left panel: \powheg,
  right panel: \mg.\label{fig:pthh_masseffects}}
\end{figure}

Because of the fact that for the $\pthh$ distribution, the tail is
predicted at the first non-trivial order, the effect of the shower on
this distribution is rather large, exceeding a factor of two beyond
$\pthh\sim 300$\,GeV, as shown in Fig.~\ref{fig:pthh_masseffects}.
However, as can also be seen from Fig.~\ref{fig:pthh_masseffects}, the
differences due to the shower are still much smaller than the
difference between the full calculation and the Born-improved HEFT
approximation, which is off by an order of magnitude for $\pthh >
500$\,GeV.  Fig.~\ref{fig:pthh_masseffects} also shows that \ftapprox
does a good job for this observable, as the tail of the $\pthh$ distribution
is determined by the real radiation. In the \powheg case, the \ftapprox curve still lies
above the full result  because the differences in the
virtual part  enter the $\bar{B}$ function in \powheg{}, which
determines the overall normalisation for the shower.

\begin{figure}
\centering
\begin{subfigure}{0.49\textwidth}
\includegraphics[width=\textwidth]{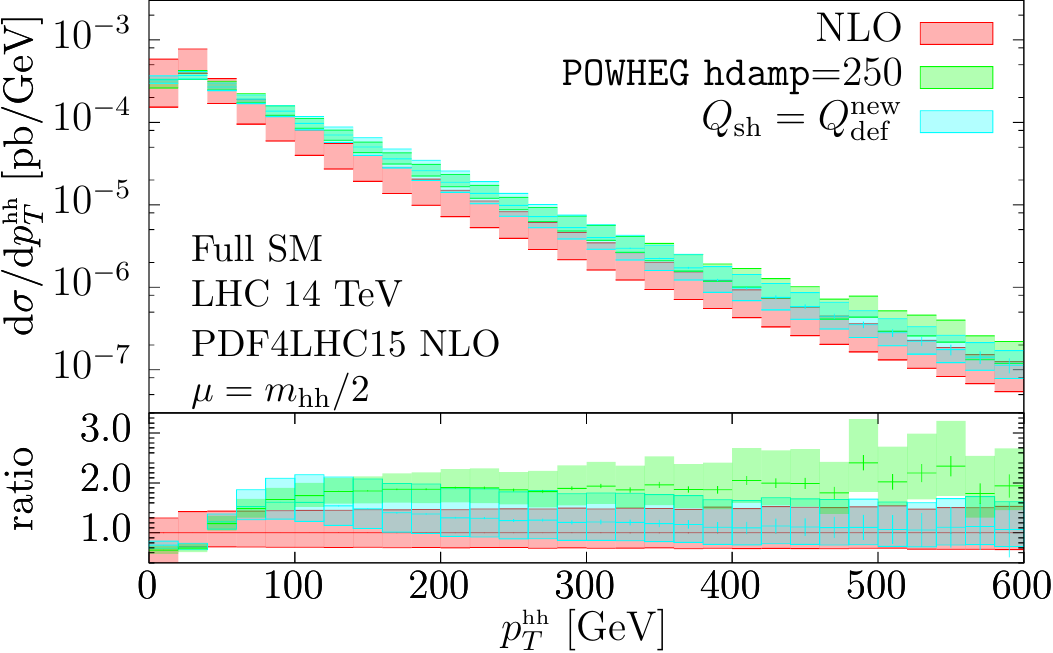}
\end{subfigure}
\begin{subfigure}{0.49\textwidth}
\includegraphics[width=\textwidth]{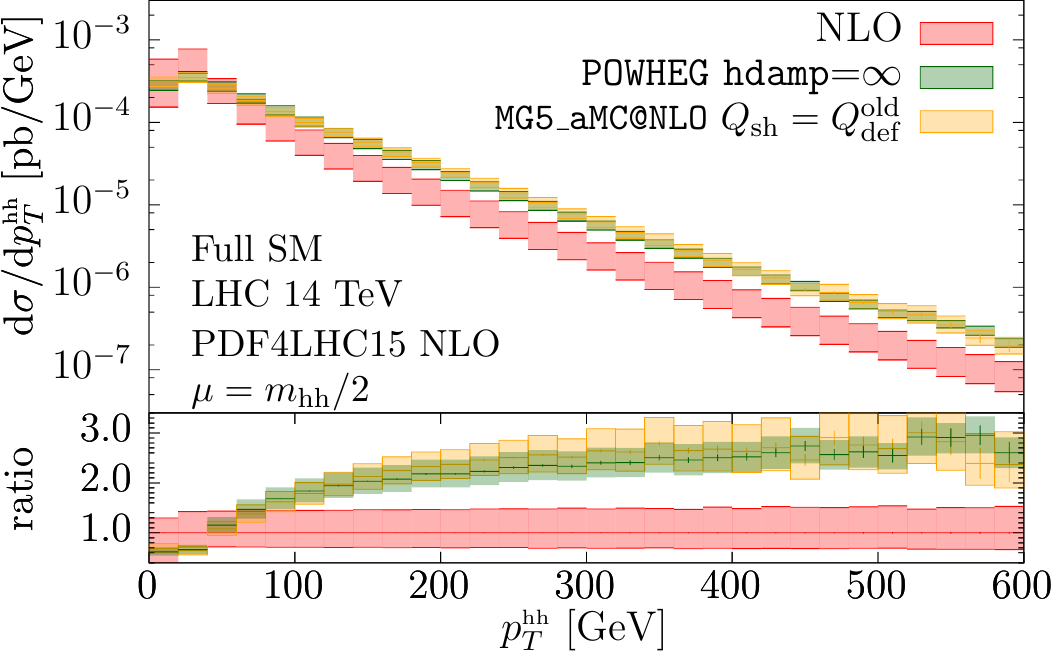}
\end{subfigure}
\caption{
$\pthh$  distribution comparing  showered results with different values for {\tt hdamp} in \powheg
  resp. the shower starting scale $Q_{\rm{sh}}$ in \mg{} compared to the fixed order result.\label{fig:shower_vs_FO}}
\end{figure}

Finally, we compare in Fig.~\ref{fig:shower_vs_FO} the fixed order
result to showered results using different values for {\tt hdamp} in
\powheg  and for the shower starting scale  $Q_{\rm{sh}}$ in \mg.
The new shower starting scale in \mg is picked in some interval with
$H_T/2$ as its maximum as stated above, while the old shower starting scale was
picked in the interval  $[0.1\,\sqrt{\hat{s}},\sqrt{\hat{s}}]$.  One can observe that with
the new shower starting scale in \mg,
$Q_{\rm{sh}}=Q_{\rm{def}}^{\rm{new}}$, the showered results match onto
the fixed order curve at large values of $\pthh$, while the latter is
not the case for \powheg with {\tt hdamp=}$\infty$ and \mg with the
old default shower starting scale.

